\documentclass[draftcls,onecolumn,12pt]{IEEEtran}
\usepackage{bm}
\usepackage{epsfig}
\usepackage{stfloats}
\usepackage{graphicx}
\usepackage{subfigure}
\usepackage{epstopdf}
\usepackage{multirow}
\usepackage[sort, compress]{cite}
\usepackage{epsfig,graphics,subfigure}
\usepackage[cmex10]{amsmath}

\usepackage{array}
\usepackage{tabulary}
\usepackage{booktabs}
\usepackage{amsfonts}
\usepackage{algorithmic}
\usepackage{algorithm}
\usepackage{array}
\usepackage{tabulary}
\usepackage{booktabs}
\usepackage{amsfonts}

\usepackage{algorithmic}
\usepackage{algorithm}

\usepackage{epstopdf}
\def\BibTeX{{\rm B\kern-.05em{\sc i\kern-.025em b}\kern-.08em
    T\kern-.1667em\lower.7ex\hbox{E}\kern-.125emX}}

\usepackage{amssymb}

\usepackage{textcomp}
\usepackage{xcolor}

\begin{document}

\title{A Cooperative Deception Strategy for Covert Communication in Presence of a Multi-antenna Adversary }

\author{Jiangbo~Si,~\IEEEmembership{Senior~Member,~IEEE},
	Zizhen~Liu,
	Zan~Li,~\IEEEmembership{Senior~Member,~IEEE},
	Hang~Hu,
    Lei~Guan,
    Chao~Wang,
	%John~Doe,~\IEEEmembership{Fellow,~OSA,}
	and~Naofal~Al-Dhahir,~\IEEEmembership{Fellow,~IEEE}}
        % <-this % stops a space
%\thanks{This paper was produced by the IEEE Publication Technology Group. They are in Piscataway, NJ.}% <-this % stops a space
%\thanks{Manuscript received April 19, 2021; revised August 16, 2021.}



\maketitle

\begin{abstract}
Covert transmission is investigated for a cooperative deception strategy, where a cooperative jammer (Jammer) tries to attract a multi-antenna adversary (Willie) and degrade the adversary's reception ability for the signal from a transmitter (Alice). For this strategy, we formulate an optimization problem to maximize the covert rate when three different types of channel state information (CSI) are available. The total power is optimally allocated between Alice and Jammer subject to Kullback-Leibler (KL) divergence constraint. Different from the existing literature, in our proposed strategy, we also determine the optimal transmission power at the jammer when Alice is silent, while existing works always assume that the jammer's power is fixed. Specifically, we apply the S-procedure to convert infinite constraints into linear-matrix-inequalities (LMI) constraints. When statistical CSI at Willie is available, we convert double integration to single integration using asymptotic approximation and substitution method. In addition, the transmission strategy without jammer deception is studied as a benchmark. Finally, our simulation results show that for the proposed strategy, the covert rate is increased with the number of antennas at Willie. Moreover, compared to the benchmark, our proposed strategy is more robust in face of imperfect CSI.
\end{abstract}

\begin{IEEEkeywords}
Cooperative deception, covert transmission, CSI, multi-antenna adversary, power allocation 	
\end{IEEEkeywords}

\section{Introduction}
\IEEEPARstart{C}{overt} transmission is an emerging high-security communication technique to transmit private information without being discovered by an adversary, which is different from traditional communication security techniques and physical layer security techniques \cite{8438892}\cite{9205225}. Traditional communication security relies on encryption systems. However, with the continuous advancement of technology, encrypted information may be deciphered by adversaries, and there is a risk of information leakage. Physical layer security relies on the characteristics of the wireless channel to realize secure transmission \cite{7514758}. However, the communication link can be exposed to the eavesdroppers. Therefore, covert communication techniques are applied in  high-security scenarios such as finance, national security, and the military. Moreover, as users become more concerned about privacy, both industry and academia are paying more attention to covert communication \cite{8714018}.
\par
%[1]Y. Zhang, Y. Shen, X. Jiang, and S. Kasahara, ¡°Mode selection and spectrum partition for D2D inband communications: A physical layer security perspective,¡± IEEE Trans. Commun., vol. 67, no. 1, pp. 623¨C638, Jan. 2019.
%[2]M. Letafati, A. Kuhestani, K. K. Wong, and M. J. Piran, ¡°A lightweight secure and resilient transmission scheme for the Internet of Things in the presence of a hostile jammer,¡± IEEE Internet Things J., vol. 8, no. 6, pp. 4373¨C4388, Mar. 2021, doi: 10.1109/JIOT.2020.3026475.
%[3]Huang Y,Wang J,Zhong C,et al.Secure Transmission in Cooperative Relaying Networks With Multiple Antennas[J]. IEEE Transactions on Wireless Communications, 2016,15(10):6843-6856.
%[4] S. Yan, Y. Cong, S. V. Hanly, and X. Zhou, ¡°Gaussian signalling for covert communications,¡± IEEE Trans. Wirel. Commun., vol. 18, no. 7, pp. 3542¨C3553, 2019
%%%%%%%%%%%% ?????????????? %%%%%%%%%%%
In the previous research on covert communication, researchers studied the performance of covert communication for various channel models, and proved the existence of the Square Root Law (SRL), i.e.,$ \underset{n\rightarrow \infty}{\lim }\frac{\mathcal{O} \left (\sqrt{n}  \right )}{n}=0  $ \cite{6584948, 7407378, 7084182, 7447769}, which was first discovered in the AWGN channel by Bash et al.\cite{6584948}. To increase the covert transmission capacity, many researchers exploited the uncertainties of the system in order to reduce the correct detection probability of the adversary. These uncertainties include noise, channel, power, and transmission slot uncertainty \cite{7805182, 7352320, 8761935, 8471218,6970786, 8379465, 7579596}. The authors proved that when noise uncertainty obeys a certain distribution, the covert communication rate can be improved \cite{7805182}\cite{7352320}. The influence of channel uncertainty was studied in the Binary Symmetric Channel (BSC), and it also helped to increase the covert rate \cite{6970786}. Additionally, a random transmit power was able to enhance covert communications performance \cite{8379465}. Moreover, the authors showed that a positive covert rate was realized with the help of Alice's transmission slot uncertainty \cite{7579596}.\par
Since the uncertainties of the channel itself were relatively limited and hard to control, researchers further used jamming or relaying nodes to transmit artificial noise (AN) for increasing the noise uncertainty and obtaining better transmission performance \cite{7964713,7421206,8445707,9456866,8519751}. A friendly uniformed jammer generated AN to help the covert transmission between Alice and Bob in \cite{7964713}\cite{7421206}. A multiple jammers scheme was also considered in \cite{8445707}, where multiple jammers cooperated to generate AN for covert transmission. In addition, for the two-hop wireless relaying systems, a full-duplex relay transmitted the jamming signal towards the adversary \cite{9456866}. To reduce the hardware, full-duplex technology was also applied at Bob, who transmitted AN to the adversary and received the confidential message from Alice simultaneously \cite{8519751}.\par
%[14] T. V. Sobers, B. A. Bash, S. Guha, et al. Covert communication in the presence of an uninformed jammer[J]. IEEE Transactions on Wireless Communications, 2017, 16(9): 6193-6206
%[15] T. V. Sobers, B. A. Bash, D. Goeckel, S. Guha, and D. Towsley, ¡°Covert communication with the help of an uninformed jammer achieves positive rate,¡± in Proc. 49th Asilomar Conf. Signals, Syst. Comput., Nov. 2015, pp. 625¨C629.
%[16] R. Soltani, D. Goeckel, D. Towsley, B. A. Bash and S. Guha. Covert Wireless Communication With Artificial Noise Generation[J]. IEEE Transactions on Wireless Communications, 2018, 17(11): 7252-7267.
%[17] Covert Rate Maximization in Wireless Full-Duplex Relaying Systems With Power Control
%[18] K. Shahzad, X. Zhou, S. Yan, et al. Achieving covert wireless communications using a full-duplex receiver[J]. IEEE Transactions on Wireless Communications, 2018, 17(12): 8517-8530
%%%%%%%%%%%% ??CSI????? %%%%%%%%%%%
Most existing works investigate covert transmission under the assumption that the  channel state information (CSI) of all links is perfectly known at Alice \cite{6584948, 7805182, 7352320,6970786, 8379465, 7579596,7964713,7421206,8445707,9456866,8519751}. However, due to the channel estimation error and the hostility of an adversary, it is challenging for Alice to obtain all links' instantaneous CSI. Thus, a few papers studied the covert transmission performance under imperfect CSI and statistical CSI. Several schemes were proposed to improve the covert transmission performance in \cite{9398675,9685518,9390203,9438645}. With imperfect CSI and instantaneous CSI, optimal beamforming and power allocation schemes were investigated in \cite{9398675}, where a regular user was deployed to cover Alice's covert transmission. For the imperfect CSI case, a robust beamforming scheme was proposed to maximize the covert rate over a multiple-input single-output (MISO) system \cite{9685518}. With the statistical CSI at the adversary, the authors of \cite{9390203} found that the performance of covert transmission improved significantly with the help of an Intelligent Reflecting Surface (IRS). Also, the performance of covert transmission assisted by IRS was studied under both imperfect CSI and statistical CSI in \cite{9438645}. \par
%[19] Strategies in Covert Communication with Imperfect Channel State Information
%[20]Intelligent Reflecting Surface (IRS)-Aided Covert Communication With Adversary¡¯s Statistical CSI
%[21]Covert Transmission Assisted by Intelligent Reflecting Surface
%%%%%%%%%%%%??????????%%%%%%%%%%%
With the maturity of multi-antenna technology, several works have focused on the covert transmission performance for multi-antenna devices, such as \cite{9398675,9381893,8878022}. Deploying multiple antennas at Alice, an increase in the covert rate was achieved with increasing the number of antennas \cite{9398675}. When a cooperative jammer was equipped with multiple antennas, beamforming was used in \cite{9381893} to maximize the covert transmission performance. However, to the best of our knowledge, few works investigated the covert performance when multiple antennas were deployed at the adversary. It is shown in \cite{8878022} that a slight increase in the number of the adversary's antennas can significantly reduce the covert rate. However, this work did not consider the jammer's effect on the covert performance and the cooperation between the jammer and Alice is neglected.\par
% [22] Robust Beamforming Design for Covert Communications
% [23] O. Shmuel, A. Cohen, O. Gurewitz, et al. Multi-antenna jamming in covert communication[C]. IEEE International Symposium on Information Theory, 2019, 987-991
% [24] Covert Wireless Communication in Presence of a Multi-Antenna Adversary and Delay Constraints
Motivated by this background, to improve the covert rate in presence of a multi-antenna adversary, we propose a novel cooperation deception strategy  where Alice and the jammer cooperate in deceiving the multi-antenna adversary. Specifically, while Alice is silent, the jammer injects AN. When Alice transmits the message covertly, the jammer adjusts the transmission power to attract the adversary's attention and cover Alice's transmission. The transmission power at Alice and the jammer are optimally allocated to maximize the covert rate with three different kinds of CSI. To further explore the pros and cons of our proposed strategy, we also study the strategy where the jammer does not deceive the adversary as a benchmark. The main contributions of this paper are summarized as follows:
\begin{itemize}
	\item We propose a cooperative deception strategy for covert transmission in presence of a multi-antenna adversary, where a jammer is used to attract the adversary's attention and cover Alice's transmission. As a result, the adversary receives the jammer's signal using the maximal ratio combining (MRC) scheme, while it receives the signal from Alice with the random combining scheme. To the best of our knowledge, this is the first work that uses a deception strategy to effectively improve the covert rate, filling the gap in previous multi-antenna Willie research.
	\item For our proposed strategy, the optimization problems are formulated to maximize the covert rate under three kinds of CSI, i.e., instantaneous CSI, imperfect CSI, and statistical CSI. Different from previous works, we not only optimize the power allocation between the jammer and Alice under the hypothesis ${\mathcal{H}_1}$, but also optimally allocate the transmission power at the jammer under hypothesis ${\mathcal{H}_0}$ simultaneously. ${\mathcal{H}_0}$ and ${\mathcal{H}_1}$ denote the hypotheses that Alice is silent or not, respectively.
	\item We adopt different methods to solve the optimization problems. For the instantaneous CSI case,  perfect covert transmission can be achieved, which means that the detection error probability is one. Therefore, we maximize the covert rate subject to equality and power constraints. For the imperfect CSI case, we consider the impact of channel estimation errors on both the Alice-to-adversary and jammer-to-adversary links. Since the existence of error introduces infinite constraints, the optimization problem becomes non-convex. To solve this problem, the S-procedure is used to transform the constraints into linear-matrix-inequalities (LMI). Furthermore, for the statistical CSI case, we derive the statistical constraint expression with the help of asymptotic approximation and the substitution method.
	\item {To gain a deep insight into the performance of our proposed strategy, the covert rate without jammer deception is studied as a benchmark. Numerical results show an interesting conclusion that for the proposed strategy, the number of antennas at adversary plays a positive effect on the covert rate. Compared to the benchmark, our proposed strategy can achieve a higher covert rate when the number of antennas at Willie is large or the channel gain of the jammer-to-adversary link is much better than that of the Alice-to-adversary link. Moreover, with severe channel estimation errors, our proposed strategy is more robust than the benchmark. Under statistical CSI, the deception strategy outperforms the benchmark even with a small number of antennas at Willie.}
\end{itemize}\par
The remainder of this paper is organized as follows. Section \uppercase\expandafter{\romannumeral2} describes the system model. The cooperative deception strategy and the covert constraint are illustrated in detail. In Section \uppercase\expandafter{\romannumeral3}, the covert rate with cooperative deception strategy is investigated under three different kinds of CSI. A special case is studied as a benchmark in Section \uppercase\expandafter{\romannumeral4}. Numerical results are presented and discussed in Section \uppercase\expandafter{\romannumeral5}. Finally, we conclude this paper in Section \uppercase\expandafter{\romannumeral6}.\par
\textit{Notations}- $(\cdot)^H$ denotes the conjugate transpose; The operator $ \left| \cdot \right|$ denotes the absolute value; $ \left\| \cdot \right\|$ denotes the Frobenius norm; $\mathrm{Re}(\cdot)$ denote the real part of a complex random variable (RV); $\exp(\sigma^2)$ denotes exponential distribution having mean $\sigma^2$; $ \mathrm{arg}(x) $ denotes the argument of $x$; $\mathbf{A}\succeq \mathbf{0}$ denotes that $\mathbf{A}$ is positive semi-definite; $f_1\left(x \right)  \Rightarrow f_2\left(x \right)$ denotes $f_1\left(x \right)$ is a sufficient condition for $f_2\left(x \right)$; $ \mathbb{C}^{m\times n}  $ denotes the $m \times n$ complex number domain; $\mathbf{I}_M$ denotes the $M \times M$ identity matrix; $ \mathcal{CN}\left ( \mu,\sigma^2 \right ) $ denotes a complex-valued circularly symmetric Gaussian distribution with mean $ \mu $ and variance $ \sigma^2 $.
\begin{table}[!t]
	\renewcommand{\arraystretch}{1.5}
	\caption{SUMMARY OF Key Variables \label{tab:table1}}
	\centering
	\resizebox{\linewidth}{!}
{
	\begin{tabular}{|c|c|}
		\hline
		Notation & Description\\
		\hline
		$\mathbf{h}_{iw}$ & Instantaneous CSI between node $i$ and Willie, $i\in \left \{ \mathrm{Alice}\left ( a \right ), \mathrm{Jammer}\left (j  \right ) \right \}$\\
		\hline
		$\Delta \mathbf{h}_{iw}$ & Estimation error of $\mathbf{h}_{iw}$, $i\in \left \{ \mathrm{Alice}\left ( a \right ), \mathrm{Jammer}\left (j  \right ) \right \}$\\
		\hline
		$\hat{\mathbf{h}} _{iw}$ & Estimated CSI between node $i$ and Willie, $i\in \left \{ \mathrm{Alice}\left ( a \right ), \mathrm{Jammer}\left (j  \right ) \right \}$\\
		\hline
		$\mathbf{u}_{w}$ & MRC weight vector of Willie\\
		\hline
		$h_{ib}$ & Instantaneous CSI between node $i$ and Bob, $i\in \left \{ \mathrm{Alice}\left ( a \right ), \mathrm{Jammer}\left (j  \right ) \right \}$\\
		\hline
		$\sigma^2_{i}$ & The noise covariance at node $i$, $i\in \left \{ \mathrm{Bob}\left ( b \right ), \mathrm{Willie}\left (w  \right ) \right \}$\\
		\hline
		$P_{\max} $ & The transmission power constraint at Alice and Jammer\\
		\hline
		$P_{a} $ & The transmission power at Alice under $\mathcal{H}_1$\\
		\hline
		$P_{j,i} $ & The transmission power constraint at Jammer under $\mathcal{H}_i$, $i\in \left \{ 0, 1 \right \}$\\
		\hline
		$\mathbb{P}_i $ & The probability of $\mathcal{H}_i$, $i\in \left \{ 0, 1 \right \}$\\
		\hline
		$\xi $ & The detection error probability at Willie\\
		\hline
		$\varepsilon $ & A required threshold of detection error probability for covert transmission \\
		\hline
		$\bar{a},\bar{b} $ & The root of $\mathrm{ln}x+\frac{1}{x}-1=\frac{2\varepsilon ^2}{L},x>0$.\\
		\hline
		$v_{aw}^2 $ & The norm-bound of channel estimation errors between Alice and Willie\\
		\hline
		$u_{jw}^2 $ & The norm-bound of channel estimation errors between Jammer and Willie\\
		\hline
		$\alpha $ & Ratio of $P_{a}$ to $P_{\max}$\\
		\hline
		$\beta $ & Ratio of $P_{j,0}$ to $P_{\max}$\\
		\hline
		$\gamma $ & Ratio of $P_{j,1}$ to $P_{\max}$\\
		\hline
	\end{tabular}
}
\end{table}

\section{System Model}
\begin{figure}[!t]
	\centering
	\includegraphics[width=2.8in]{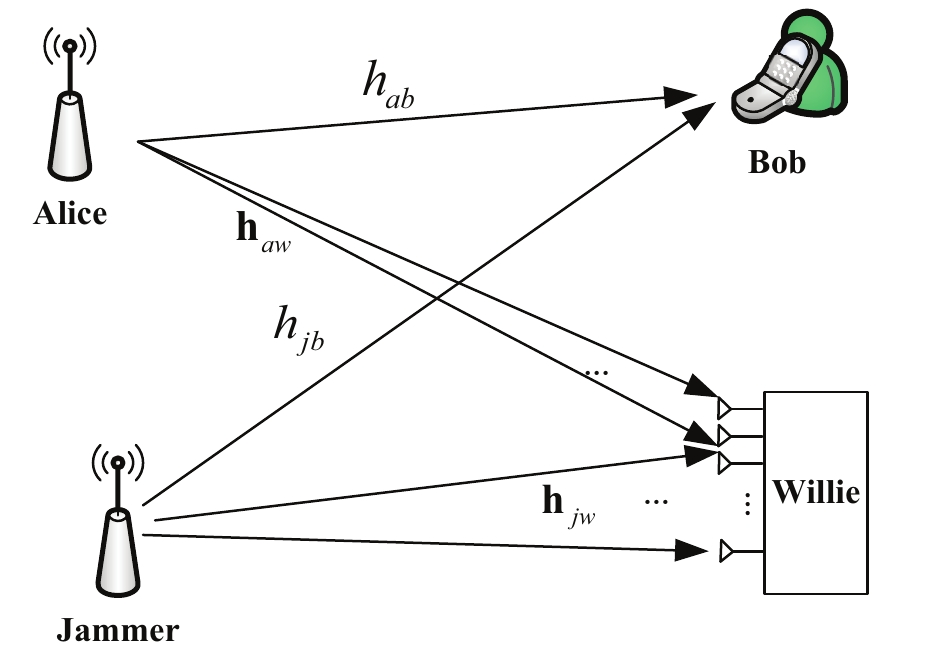}
	\caption{System model}
	\label{fig_1}
\end{figure}

As shown in Fig. \ref{fig_1}, Alice tries to transmit messages to Bob without being detected by an adversary (Willie), who tries to detect whether there exists a transmission from Alice or not. Moreover, Jammer, Alice's coordinator, tries to confuse Willie and helps to cover Alice's transmission.\par
In this model, we consider a scenario where Willie uses $M > 1$ antennas for detection, whereas Alice, Jammer and Bob are equipped
with a single antenna. Each channel is assumed to follow the Rayleigh fading model. The instantaneous CSI of the Alice $\rightarrow  $ Bob and Jammer $\rightarrow  $ Bob links are, respectively, denoted by $h_{ab} \sim \mathcal{CN}\left ( 0,\sigma _{ab}^{2} \right )$ and $h_{jb} \sim \mathcal{CN}\left ( 0,\sigma _{jb}^{2} \right )$. In addition, $\textbf{h}_{aw}\in \mathbb{C}^{M\times 1} \sim \mathcal{CN}\left ( \mathbf{0},\sigma _{aw}^{2} \mathbf{I}\right )$ and $\textbf{h}_{jw}\in \mathbb{C}^{M\times 1} \sim \mathcal{CN}\left ( \mathbf{0},\sigma _{jw}^{2} \mathbf{I}\right )$ denote the instantaneous CSI of the Alice $\rightarrow  $ Willie and Jammer $\rightarrow  $ Willie links, respectively. The total maximum transmission power is $P_{\max}$, and the power can be allocated to Alice and Jammer due to their coordinated transmission. In addition, the key variables are given in Table \uppercase\expandafter{\romannumeral1}.\par
\subsection{Cooperative Deception Strategy}
We assume that Alice transmits $L$ symbols in a transmission time slot with the symbol index of $l$, and Jammer injects AN at each time slot. Let $\mathcal{H}_0$ denotes the hypothesis that Alice does not transmit private information to Bob, while $\mathcal{H}_1$ denotes the hypothesis that Alice transmits the confidential messages to Bob \cite{2005Testing}. From Willie's perspective, the transmitted signals from Alice and Jammer are given by
%[25]E. L. Lehmann and J. P. Romano, Testing Statistical Hypotheses, Springer New York, 2005
\begin{equation}	
	{x_{a}^{l}} = \begin{cases}
		0,&{{\mathcal{H}_0}} \\
		{x_{b}^{l},}&{{\mathcal{H}_1}}
	\end{cases}
	\quad , and \quad \quad {x_{j}^{l}} = \begin{cases}
		{x_{j,0}^{l},}&{{\mathcal{H}_0}} \\
		{x_{j,1}^{l},}&{{\mathcal{H}_1}}
	\end{cases},
\end{equation}
where $x_{b}^{l}$, $x_{j,0}^{l}$ and $x_{j,1}^{l}$ are transmitted information with unity power from Alice, and Jammer under ${{\mathcal{H}_0}}$ and  ${{\mathcal{H}_1}}$, respectively. As mentioned before, the transmit power satisfies $P_a+P_{j,1} \leq P_{\max}$ under $\mathcal{H}_1$, and also $P_{j,0} \leq P_{\max}$ under $\mathcal{H}_0$.Thus, the received signal at Bob is given by
\begin{equation}
	{y_{b}^{l}} = \begin{cases}
		\sqrt{P_{j,0}}h_{jb}x_{j,0}^{l}+n_b^{l},&{{\mathcal{H}_0}} \\
		{\sqrt{P_{a}}h_{ab}x_{b}^{l}+\sqrt{P_{j,1}}h_{jb}x_{j,1}^{l}+n_b^{l},}&{{\mathcal{H}_1}}
	\end{cases}
\end{equation}
where $n_b^{l}\sim \mathcal{CN}\left (0,\sigma _{b}^{2} \right )$ is the noise at Bob. Correspondingly, the signal received at Willie can be written as
\begin{equation}
	{\mathbf{y}_{w}^{l}} = \begin{cases}
		\sqrt{P_{j,0}}\mathbf{h}_{jw}x_{j,0}^{l}+\mathbf{n}_w^{l},&{{\mathcal{H}_0}} \\
		{\sqrt{P_{a}}\mathbf{h}_{aw}x_{b}^{l}+\sqrt{P_{j,1}}\mathbf{h}_{jw}x_{j,1}^{l}+\mathbf{n}_w^{l},}&{{\mathcal{H}_1}}
	\end{cases}
\end{equation}
where $\mathbf{n}_w^{l}\sim \mathcal{CN}\left (0,\sigma _{w}^{2}\mathbf{I}\right )$ is the background noise at Willie.
Since Jammer transmits the signal, we assume that Willie has the ability to estimate the channel state information with blind signal processing, and maximum ratio combining (MRC) is applied to Willie for the diversity gain. To explore the advantages of the cooperative deception strategy in this scenario, we assume that Jammer already transmitted an interference signal before Alice transmitted the confidential messages. Therefore, Willie is more likely to use MRC for Jammer than Alice. Thus, the MRC weight vector is given by $\mathbf{u}_w=\frac{\mathbf{h}_{jw}^{H}}{\left \|\mathbf{h}_{jw}  \right \|}$, and the actual received signal at Willie after using MRC is given by
\begin{equation}\label{eq:4}
\begin{aligned}
&{\hat{y}_{w}^{l}} \\
=&\mathbf{u}_w\mathbf{y}_{w}^{l}\\
	=& \begin{cases}
	\sqrt{P_{j,0}}\left \|\mathbf{h}_{jw}  \right \|x_{j,0}^{l}+\frac{\mathbf{h}_{jw}^{H}}{\left \|\mathbf{h}_{jw}  \right \|}\mathbf{n}_w^{l},&{{\mathcal{H}_0}} \\
	{\sqrt{P_{a}}\frac{\mathbf{h}_{jw}^{H}}{\left \|\mathbf{h}_{jw}  \right \|}\mathbf{h}_{aw}x_{b}^{l}+\sqrt{P_{j,1}}\left \|\mathbf{h}_{jw}  \right \|x_{j,1}^{l}+\frac{\mathbf{h}_{jw}^{H}}{\left \|\mathbf{h}_{jw}  \right \|}\mathbf{n}_w^{l}.}&{{\mathcal{H}_1}}
\end{cases}
\end{aligned}
\end{equation}
When Willie uses MRC to receive the signal from Jammer, the signal from Alice is received by Willie with random combining \cite{9122034}. As a result, the multi-antenna diversity gain at Willie cannot be used to receive the signal from Alice, and the covert rate can be improved. Under the help of \eqref{eq:4}, the average power of the received signal at Willie is given by
\begin{equation}\label{eq:5}
		\left| \hat{y}_{w}^{l}\right| ^2
		= \begin{cases}
			P_{j,0}\left \|\mathbf{h}_{jw}  \right \|^2+\sigma_{w}^2,&{{\mathcal{H}_0}} \\
			P_{a}\left |h_{aw}  \right |^2+P_{j,1}\left \|\mathbf{h}_{jw}  \right \|^2+\sigma_{w}^2, &{{\mathcal{H}_1}}
		\end{cases}
\end{equation}
where $h_{aw}$ is the element of $\mathbf{h}_{aw}$.
\par
\subsection{Covert Constraint}
In general, the prior probabilities of hypotheses ${\mathcal{H}_0}$ and ${\mathcal{H}_1}$ are assumed to be equal \cite{7964713}. Hence, the detection error probability (DEP) $\xi $ can be expressed as
\begin{equation}\label{eq:6}
	\xi = P_{MD}+P_{FA} = \Pr\left( {\mathcal{D}_0}|{\mathcal{H}_1}\right) +\Pr\left( {\mathcal{D}_1}|{\mathcal{H}_0}\right),
\end{equation}
%[14]T. V. Sobers, B. A. Bash, S. Guha, D. Towsley, and D. Goeckel, ¡°Covert communication in the presence of an uninformed jammer,¡± IEEE Trans. Wireless Commun., vol. 16, no. 9, pp. 6193¨C6206, Sep. 2017
where $P_{MD}$ and $P_{FA}$ denote, respectively, the miss detection probability and the false alarm probability. $\mathcal{D}_1$ and $\mathcal{D}_0$ indicate whether the transmission from Alice to Bob is present or not. When DEP is larger than a predetermined threshold $1-\varepsilon$, the transmission can be treated as covert, e.g. the covert constraint is $\xi \geq 1-\varepsilon $. With some mathematical calculations and derivations, $P_{MD}$ and $P_{FA}$ are, respectively, given as \cite{9398675}
\begin{equation}\label{eq:7}
	P_{MD}=\frac{\gamma \left ( L, \frac{\theta ^*}{\left | h_{aw} \right |^2P_a+\left \| \mathbf{h}_{jw} \right \|^2P_{j,1}+\sigma_w^2} \right )}{\Gamma \left ( L \right )},
\end{equation}
and
\begin{equation}\label{eq:8}
	P_{FA}=1-\frac{\gamma \left ( L, \frac{\theta ^*}{\left \| \mathbf{h}_{jw} \right \|^2P_{j,0}+\sigma_w^2} \right )}{\Gamma \left ( L \right )},
\end{equation}
where $\gamma \left(a,b \right) $ is the lower incomplete Gamma function, $\Gamma\left(x \right) $ is the complete Gamma function, and $\theta^*$ is the optimal threshold of Willie's detector which is given by
\begin{equation}\label{eq:9}
	\theta^* = L \frac{\lambda_0\lambda_1}{\lambda_1-\lambda_0}\ln \frac{\lambda_1}{\lambda_0},
\end{equation}
where $\lambda _0 = \left \|\mathbf{h}_{jw}  \right \|^2P_{j,0}+\sigma _{w}^{2}$ and $\lambda _1 =\left | h_{aw} \right |^2P_{a}+\left \|\mathbf{h}_{jw}  \right \|^2P_{j,1}+\sigma _{w}^{2}$. From \eqref{eq:6}-\eqref{eq:8}, we find that using $P_{MD}+P_{FA} \geq 1 - \varepsilon$ as a constraint would make the problem difficult to solve due to the special functions involved. Thus, we use another expression of DEP.\par
For the $l^{th}$ symbol, the probabilities of ${\mathcal{H}_0}$ and ${\mathcal{H}_1}$ are $\mathbb{P}_0$ and $\mathbb{P}_1$, respectively, and the corresponding probability density functions are $p_0(x)$ and $p_1(x)$. Willie detects $L$ symbols in a time slot, and each symbol is independent of the other symbols. Thus, for Willie, $P\left \{ \mathcal{H}_0 \ \mathrm{is\ true} \right \}=\left (\mathbb{P}_0  \right )^L$, $P\left \{ \mathcal{H}_1 \ \mathrm{is\ true} \right \}=\left (\mathbb{P}_1  \right )^L$. Both ${\hat{y}_{w}^{l}}$ and ${\left |\hat{y}_{w}^{l}  \right |}$ can be used for calculating $p_0(x)$ and $p_1(x)$\cite{9456866}, and the former one is used in this paper, which is expressed as
%[22]Robust Beamforming Design for Covert Communications
%[17]Covert Rate Maximization in Wireless Full-Duplex Relaying Systems With Power Control
\begin{equation}\label{eq:10}
	p_0\left ( \hat{y}_w^l \right )=\frac{1}{\pi\lambda _0}\mathrm{exp}\left ( -\frac{\left |\hat{y}_w^l  \right |^{2} }{\lambda _0} \right ),
\end{equation}
\begin{equation}\label{eq:11}
	p_1\left ( \hat{y}_w^l \right )=\frac{1}{\pi\lambda _1}\mathrm{exp}\left ( -\frac{\left |\hat{y}_w^l  \right |^{2} }{\lambda _1} \right ).
\end{equation}
Different from \eqref{eq:6}, let
\begin{equation}\label{eq:12}
	\xi = 1-V_{T}\left(\mathbb{P}_{0}^L, \mathbb{P}_{1}^L \right),
\end{equation}
where $V_{T}\left(x_1, x_2 \right)$ is the total variation between $x_{1}$ and $x_{2}$. Then, using Pinsker's inequality \cite{2006Elements}, we have
\begin{equation}\label{eq:13}
	\begin{aligned}
	&V_{T}\left(\mathbb{P}_{0}^L, \mathbb{P}_{1}^L \right) \leq \sqrt{\frac{1}{2}D\left ( \mathbb{P}_{0}^L||\mathbb{P}_{1}^L \right )} =\sqrt{\frac{L}{2}D\left ( \mathbb{P}_{0}||\mathbb{P}_{1} \right )},\\
	&V_{T}\left(\mathbb{P}_{0}^L, \mathbb{P}_{1}^L \right) \leq \sqrt{\frac{1}{2}D\left ( \mathbb{P}_{1}^L||\mathbb{P}_{0}^L \right )}=\sqrt{\frac{L}{2}D\left ( \mathbb{P}_{1}||\mathbb{P}_{0} \right )},
	\end{aligned}
\end{equation}
where $D\left ( \mathbb{P}_{0}||\mathbb{P}_{1} \right )$ is the Kullback-Leibler (KL) divergence \cite{8379465} from $p_0\left(x \right) $ to $p_1\left(x \right)$. Since the solution method under $D\left ( \mathbb{P}_{0}\|\mathbb{P}_{1} \right  )$ and $D\left ( \mathbb{P}_{1}\|\mathbb{P}_{0} \right  )$ are consistent, this paper takes $D\left ( \mathbb{P}_{0}\|\mathbb{P}_{1} \right  )$ as an example which is given by
\begin{equation}\label{eq:14}
	D\left ( \mathbb{P}_{0}||\mathbb{P}_{1} \right )=\ln \frac{\lambda_1}{\lambda_0}+\frac{\lambda_0}{\lambda_1}-1.
\end{equation}
Substituting \eqref{eq:13} into \eqref{eq:12}, the covert constraint can be expressed as
\begin{equation}\label{eq:15}
	D\left ( \mathbb{P}_{0}^{L}\|\mathbb{P}_{1}^{L} \right  )=LD\left ( \mathbb{P}_{0}\|\mathbb{P}_{1} \right  )\leq 2\varepsilon ^2.
\end{equation}
We use \eqref{eq:15} instead of \eqref{eq:6} to express the covert constraint due to low complexity.
%[26] T. M. Cover and J. A. Thomas, Elements of Information Theory, New York:Wiley, 2006.
%[12]S. Yan, B. He, X. Zhou, Y. Cong, and A. L. Swindlehurst, ¡°Delay-intolerant covert communications with either fixed or random transmit power,¡± IEEE Trans. Inf. Forensics Secur., vol. 14, no. 1, pp. 129¨C140, Jan. 2019.
\subsection{Different CSI scenarios}
We assume that due to the cooperative relationship between Alice, Jammer, and Bob, Alice and Jammer can estimate Bob's CSI accurately. However, it is challenging for Alice to obtain the CSI of Willie's links if Willie is silent and unfriendly to Alice. Thus, we consider the following three scenarios:
\subsubsection{Scenario 1 (instantaneous CSI)}
As assumed in many papers, Willie is a legitimate user to Alice and Jammer, while it is hostile to Bob. In this case, Alice and Jammer know instantaneous CSI of the Alice-to-Willie link and the Jammer-to-Willie link and use it to assist the covert transmission from Alice. Moreover, the covert rate with instantaneous CSI can be an upper bound over that with imperfect CSI and statistical CSI. Hence, it is necessary to investigate the covert rate under this scenario.
\subsubsection{Scenario 2 (imperfect CSI)}
Consider a more realistic scenario where Willie has only limited cooperation with Alice and Jammer. In this case, we assume that the CSI of Alice-to-Willie link and Jammer-to-Willie link are imperfect. More specifically, in the presence of channel estimation errors, the imperfect CSI of Willie's links can be modeled as
\begin{equation}\label{eq:16}
	\mathbf{h}_{aw}=\mathbf{\hat{h}}_{aw}+\Delta \mathbf{h}_{aw}, \quad and \and \quad \mathbf{h}_{jw}=\mathbf{\hat{h}}_{jw}+\Delta \mathbf{h}_{jw},
\end{equation}
where $\mathbf{\hat{h}}_{aw}$ and $\mathbf{\hat{h}}_{jw}$ are the estimated CSI of Alice-to-Willie link and Jammer-to-Willie link, respectively. The corresponding CSI error vectors are $\Delta \mathbf{h}_{aw}$ and $\Delta \mathbf{h}_{jw}$, which are characterized by the norm-bounded model, i.e.,
\begin{equation}\label{eq:17}
\left\| \Delta \mathbf{h}_{aw} \right\|^2  \leq v_{aw}^2, \quad and \quad \left\| \Delta \mathbf{h}_{jw} \right\|^2 \leq u_{jw}^2,
\end{equation}
where $v_{aw}^2>0$ and $u_{jw}^2>0$.
\subsubsection{Scenario 3 (statistical CSI)}
Further, assuming that Willie is an enemy adversary and deliberately hides himself, then Alice and Jammer only know the statistical CSI for Willie's links, i.e., variances $\sigma _{aw}^{2}$ and $\sigma _{jw}^{2}$.

\section{Optimal Power Allocation for Different CSI}
In this section, we investigate the optimal power allocation at Alice and Jammer for different CSI scenarios where the cooperative deception strategy is applied, and Jammer transmits AN to attract Willie's attention. As a result, Willie uses MRC to combine the signal from Jammer, and can only use random combining scheme to receive the signal from Alice. Thus, the advantage of multiple antennas cannot be used to detect Alice's transmission, and Alice's signal is received with a single antenna gain. In the following, we discuss the optimal power allocation for three different CSI.

\subsection{Instantaneous CSI Scenario}
In this section, we maximize the covert rate by optimizing the power allocation at Alice and Jammer under ${\mathcal{H}_0}$ and ${\mathcal{H}_1}$ when all links' instantaneous CSI are available. Specifically, for the proposed strategy, the covert rate $R_{b}$ is maximized subject to the perfect covert and total power constraints, and the optimization problem can be formulated as
\begin{subequations}\label{eq:18}
	\begin{align}
		&\underset{P_a,P_{j,0},P_{j,1}}{\mathrm{max}}{R_{b}}, \label{eq:18a}\\
	s.t. 	\ &D\left ( \mathbb{P}_{0}\|\mathbb{P}_{1} \right  )=0, \label{eq:18b}\\
			&P_a\left | h_{aw} \right |^2\leq P_{j,1}\left \| \mathbf{h}_{jw} \right \|^2, \label{eq:18c}\\
			&P_{j,1}+P_{a} \leq P_{\mathrm{max}},\label{eq:18d}\\
			&P_{j,0} \leq P_{\mathrm{max}},\label{eq:18e}
	\end{align}
\end{subequations}
where
\begin{equation}\label{eq:19}
	{R_{b}}=\mathrm{log}\left ( 1+\frac{P_a\left | h_{ab} \right |^2}{P_{j,1}\left | h_{jb} \right |^2+\sigma _{b}^{2}} \right ).
\end{equation}
It is possible to achieve perfect covert transmission with instantaneous CSI by adjusting the transmit power. Thus \eqref{eq:18b} is an equality constraint. \eqref{eq:18c} implies that the power received by Willie from Jammer under ${\mathcal{H}_1}$ is greater than the power received from Alice, and Willie can be deceived to focusing the signal from Jammer{\footnote{In this paper, we assume that Willie can use MRC for only one user, and cannot use MRC for both Jammer and Alice simultaneously.}}. Otherwise, Willie will perceive that the power sent by Alice is much stronger and use MRC on Alice's transmission, which will make the deception invalid.\par
Notice that \eqref{eq:19} is an increasing function of $P_a$ and a decreasing function of $P_{j,1}$, combined with the total power constraint \eqref{eq:18d}, the objective function can be simplified to $P_a$. Since the only one root of $f\left(x \right) = \ln x+\frac{1}{x}-1 $ is $x = 1$, \eqref{eq:18b} holds only if $\lambda_0=\lambda_1$, i.e. \eqref{eq:18} can be restated as
\begin{subequations}\label{eq:20}
	\begin{align}
		&\underset{P_a,P_{j,0},P_{j,1}}{\mathrm{max}}P_a, \label{eq:20a}\\
s.t. 	\ & P_{a}\left | h_{aw} \right |^2=\left( P_{j,0}-P_{j,1}\right)\left \|\mathbf{h}_{jw}  \right \|^2, \label{eq:20b}\\
		&P_a\left | h_{aw} \right |^2\leq P_{j,1}\left \| \mathbf{h}_{jw} \right \|^2, \label{eq:20c}\\
		&P_{j,1}+P_{a} \leq P_{\mathrm{max}},\label{eq:20d}\\
		&P_{j,0} \leq P_{\mathrm{max}}.\label{eq:20e}
	\end{align}
\end{subequations}
Substitute \eqref{eq:20b} into \eqref{eq:20c}, we get the following inequality:
\begin{equation}\label{eq:21}
	P_{j,1}\leq P_{j,0}\leq 2P_{j,1}.
\end{equation}
To simplify the calculation, we assume that the Jammer-to-Willie link follows the same channel distribution at each antenna,  and let $ \left \|\mathbf{h}_{jw}  \right \|^2=M\left| h_{jw}\right|^2$, where $h_{jw}$ is the element of $\mathbf{h}_{jw}$. Meanwhile, from \eqref{eq:20b}, we find that maximizing $P_a$ is equivalent to maximizing $\left( P_{j,0}-P_{j,1} \right)$. Thus \eqref{eq:21} can be transformed into $P_{j,0}=2P_{j,1}$ and \eqref{eq:20b} can be transformed into
\begin{equation}\label{eq:22}
	P_{a}=\frac{P_{j,1}M\left | h_{jw} \right |^2}{\left | h_{aw} \right |^2}.
\end{equation}
\emph{Discussion:} From \eqref{eq:22}, we discuss two cases according to the relationship between $\left | h_{aw} \right |^2$ and $\left | h_{jw} \right |^2$ as follows:
\subsubsection{Case of $\left | h_{aw} \right |^2\geq \left | h_{jw} \right |^2$}
When $M\leq\frac{\left | h_{aw} \right |^2}{\left | h_{jw} \right |^2}$, $P_a$ increases with $P_{j,1}$ and $M$. Considering \eqref{eq:20e}, we can derive the optimal power at Alice and Jammer as
\begin{equation}\label{eq:23}
	P_a^*=M\frac{P_{\mathrm{max}}\left | h_{jw} \right |^2}{2\left | h_{aw} \right |^2},
	\  P_{j,1}^* = \frac{P_{\mathrm{max}}}{2}\  \mathrm{and}
	\  P_{j,0}^* = P_{\mathrm{max}}.
\end{equation}
When $M>\frac{\left | h_{aw} \right |^2}{\left | h_{jw} \right |^2}$, restricted by \eqref{eq:23}, $P_{j,1}$ cannot be increased any more. By combining $P_{j,1}+P_{a} = P_{\mathrm{max}}$ and \eqref{eq:22}, we get
\begin{equation}\label{eq:24}
	\begin{split}
	P_a^*&=\frac{P_{\mathrm{max}}M\left | h_{jw} \right |^2}{\left | h_{aw} \right |^2+M\left | h_{jw} \right |^2},
	\  P_{j,1}^* = 1-P_a^*\  \mathrm{and}\\
	  P_{j,0}^* &= 2\left(1-P_a^* \right) .
	\end{split}
\end{equation}\par
{To achieve perfect covert and the maximum covert rate, when $M\leq\frac{\left | h_{aw} \right |^2}{\left | h_{jw} \right |^2}$, $P_{j,1}$ and $P_{j,0}$ are fixed, while $P_a$ increases linearly with $M$. However, when $M>\frac{\left | h_{aw} \right |^2}{\left | h_{jw} \right |^2}$, $P_{j,0}$ and $P_{j,1}$ decrease as $P_{a}$ increases.}

\subsubsection{Case of $\left | h_{aw} \right |^2 < \left | h_{jw} \right |^2$}
Since $M\in \mathbb{N}^+$, $M>\frac{\left | h_{aw} \right |^2}{\left | h_{jw} \right |^2}$ is the only case to consider and the optimal power allocation is the same as \eqref{eq:24} that $P_{j,0}$ and $P_{j,1}$ decrease as $P_{a}$ increases.

\subsection{Imperfect CSI Scenario}
The instantaneous CSI scenario in the previous section is the ideal case. In contrast, the imperfect CSI scenario considering the channel estimation error \cite{9634882} is more general. Hence, in this section, we formulate the optimization problems under the imperfect CSI scenario. Due to the errors in channel estimation, perfect covert transmission is difficult to achieve \cite{8714018}. We used imperfect covert constraints $D\left ( \mathbb{P}_{0}\|\mathbb{P}_{1} \right  )\leq \frac{2\varepsilon ^2}{L}$ instead of the perfect one $D\left ( \mathbb{P}_{0}\|\mathbb{P}_{1} \right  )=0$, and channel bounded error models \eqref{eq:16} \eqref{eq:17} are considered in the constraints. Mathematically, the optimization problem is formulated as
%[27]Performance Analysis for Dual-Hop Covert Communication System with Outdated CSI
%[4]S. Yan, Y. Cong, S. V. Hanly, and X. Zhou, ¡°Gaussian signalling for covert communications,¡± IEEE Trans. Wireless Commun., vol. 18, no. 7, pp. 3542¨C3553, Jul. 2019
\begin{subequations}\label{eq:25}
	\begin{align}
		&\underset{P_a,P_{j,0},P_{j,1}}{\mathrm{max}}P_a, \label{eq:25a}\\
		s.t. 	\  &D\left ( \mathbb{P}_{0}\|\mathbb{P}_{1} \right  )\leq \frac{2\varepsilon ^2}{L}, \label{eq:25b}\\
		&P_a\left | h_{aw} \right |^2\leq P_{j,1}\left \| \mathbf{h}_{jw} \right \|^2, \label{eq:25c}\\
		&P_{j,1}+P_{a} \leq P_{\mathrm{max}},\label{eq:25d}\\
		&P_{j,0} \leq P_{\mathrm{max}},\label{eq:25e}\\
		&\mathbf{h}_{aw}=\mathbf{\hat{h}}_{aw}+\Delta \mathbf{h}_{aw}, \ \left\| \Delta \mathbf{h}_{aw} \right\|^2  \leq v_{aw}^2, \label{eq:25f}\\
		&\mathbf{h}_{jw}=\mathbf{\hat{h}}_{jw}+\Delta \mathbf{h}_{jw}, \ \left\| \Delta \mathbf{h}_{jw} \right\|^2  \leq u_{jw}^2. \label{eq:25g}
	\end{align}
\end{subequations}
\eqref{eq:25b} can be equivalently written as
\begin{equation}\label{eq:26}
	\bar{a}\leq \frac{P_{a}\left | h_{aw} \right |^2+P_{j,1}\left \|\mathbf{h}_{jw}  \right \|^2+\sigma _{w}^{2}}{P_{j,0}\left \|\mathbf{h}_{jw}  \right \|^2+\sigma _{w}^{2}}\leq \bar{b},
\end{equation}
where $0<\bar{a}<1$ and $\bar{b}>1$ is the root of $\mathrm{ln}x+\frac{1}{x}-1=\frac{2\varepsilon ^2}{L},x>0$.
Since Willie does not use MRC to receive Alice's signal, there is only a single antenna gain at Willie for Alice's transmit power. Therefore \eqref{eq:25f} can be written approximately as
\begin{equation}\label{eq:27}
	h_{aw}=\hat{h}_{aw}+\Delta h_{aw}, \  \left| \Delta h_{aw} \right| ^2 \leq \frac{v_{aw}^2}{M}.
\end{equation}\par
Notice that the error models in \eqref{eq:25g} and \eqref{eq:27} lead to an infinite number of constraints which makes the optimization problem intractable. Thus, the S-procedure is employed for the inequality constraints to convert non-Linear-matrix-inequalities (non-LMI) conditions into LMIs \cite{2017A}. More specifically, with the help of the S-procedure, the boundary constraints combined with \eqref{eq:25c} and \eqref{eq:26} are transformed into LMI. It is noted that \eqref{eq:26} is equivalently re-expressed as
\begin{equation}\label{eq:28}
	\begin{split}
		&\left (\bar{a}P_{j,0}-P_{j,1}  \right )\left ( \mathbf{\hat{h}}_{jw}+\Delta \mathbf{h}_{jw}  \right )^{H}\left ( \mathbf{\hat{h}}_{jw}+\Delta \mathbf{h}_{jw}  \right ) \\
		&-P_a\left ( \hat{h}_{aw}+\Delta h_{aw}  \right )^{H} \left ( \hat{h}_{aw}+\Delta h_{aw}  \right )+\left ( \bar{a}-1 \right )\sigma _w^2 \leq 0,
	\end{split}
\end{equation}
and
\begin{equation}\label{eq:29}
	\begin{split}
		&\left (P_{j,1}-\bar{b}P_{j,0}  \right )\left ( \mathbf{\hat{h}}_{jw}+\Delta \mathbf{h}_{jw}  \right )^{H}\left ( \mathbf{\hat{h}}_{jw}+\Delta \mathbf{h}_{jw}  \right )\\
		&+P_a\left ( \hat{h}_{aw}+\Delta h_{aw}  \right )^{H} \left ( \hat{h}_{aw}+\Delta h_{aw}  \right )- \left ( \bar{b}-1 \right )\sigma _w^2\leq 0.		
	\end{split}
\end{equation}
Similarly, \eqref{eq:25c} is also given as
\begin{equation}\label{eq:30}
	\begin{split}
		&-P_{j,1}\left ( \mathbf{\hat{h}}_{jw}+\Delta \mathbf{h}_{jw}  \right )^{H}\left ( \mathbf{\hat{h}}_{jw}+\Delta \mathbf{h}_{jw}  \right )\\
		&+P_a\left ( \hat{h}_{aw}+\Delta h_{aw}  \right )^{H} \left ( \hat{h}_{aw}+\Delta h_{aw}  \right )\leq 0.
	\end{split}
\end{equation}  \par
Noted that equations \eqref{eq:28}-\eqref{eq:30} all satisfy the function
\begin{equation}\label{eq:31}
	f_m(x)=x^HA_mx+2\mathrm{Re}\left \{ b_m^H x\right \}+c_m,
\end{equation}
where $m\in \left \{ 1,2,3 \right \},x\in \mathbb{C}^{\left( M+2\right) \times1}, A_m\in \mathbb{C}^{M+2},b_m\in \mathbb{C}^{\left( M+2\right) \times1}$ and $c_m\in \mathbb{R}^{1\times 1}$. Due to space limitations, we only take \eqref{eq:28} as an example, and \eqref{eq:29} \eqref{eq:30} can be calculated in the same way. For \eqref{eq:28}, these parameters are given by
\begin{equation}\label{eq:32}
	\begin{split}
	x &= \begin{bmatrix}
		\Delta \mathbf{h}_{jw} & \Delta h_{aw} & 1
	\end{bmatrix}^H, \\
	A_1&=diag\left(\left (\bar{a}P_{j,0}-P_{j,1}  \right )\cdot \mathbf{I}^M, -P_a, 0 \right),\\
	b_1&= \begin{bmatrix}
		\left (\bar{a}P_{j,0}-P_{j,1}  \right )\mathbf{\hat{h}}_{jw}& -P_a\hat{h}_{aw} & 0
	\end{bmatrix}\ \mathrm{and}\\
	c_1&=\left (\bar{a}P_{j,0}-P_{j,1}  \right )\left \|\mathbf{\hat{h}}_{jw}  \right \|^{2}-P_a\left| \hat{h}_{aw}\right| ^2+\left ( \bar{a}-1 \right )\sigma_{w}^{2}.
	\end{split}
\end{equation}
By using matrix formulation, \eqref{eq:28} can be written as
\begin{equation}\label{eq:33}
\begin{bmatrix}
		\left (\bar{a}P_{j,0}-P_{j,1}  \right )\mathbf{I}_{M} & \mathbf{0}  & \left (\bar{a}P_{j,0}-P_{j,1}  \right )\mathbf{\hat{h}}_{jw}\\
		\mathbf{0} &  -P_a & -P_a\hat{h}_{aw}\\
		\left (\bar{a}P_{j,0}-P_{j,1}  \right )\mathbf{\hat{h}}_{jw}^H & -P_a\hat{h}_{aw}^H &  c_1\\
\end{bmatrix} \preceq  \textbf{0}.
\end{equation}
Similarly, \eqref{eq:29} and \eqref{eq:30} can be transformed into formulations similar to \eqref{eq:33},  and the error bound constraints can also be transformed as
\begin{equation}\label{eq:34}
	diag\left(\mathbf{I}_M, 1, -u_{jw}^2-\frac{v_{aw}^2}{M}  \right)\preceq  \textbf{0}.
\end{equation}\par
Then, according to the S-procedure[9], the implication \eqref{eq:34} $\Rightarrow $ \eqref{eq:33}, \eqref{eq:34}  $\Rightarrow $ \eqref{eq:29} and \eqref{eq:34} $\Rightarrow $ \eqref{eq:30} hold if and only if there exist variables $\eta_1>0$, $\eta_2>0$ and $\eta_3>0$, respectively, such that
\begin{equation}\label{eq:35}
\begin{bmatrix}
	\left ( \eta_1-\varsigma  \right )\mathbf{I}_M & \mathbf{0} &-\varsigma \mathbf{\hat{h}}_{jw} \\
	\mathbf{0} & \eta_1+P_{a}& P_{a}\hat{h}_{aw} & \\
	-\varsigma \mathbf{\hat{h}}_{jw}^H & P_{a}\hat{h}_{aw} ^H & \eta_1\left ( -u_{jw}^2-\frac{v_{aw}^2}{M} \right )-c_1
\end{bmatrix}\succeq \mathbf{0},
\end{equation}
where $\varsigma = \bar{a}P_{j,0}-P_{j,1}$.
\begin{equation}\label{eq:36}
	\begin{bmatrix}
		\left ( \eta_2-\zeta  \right )\mathbf{I}_M & \mathbf{0} &-\zeta \mathbf{\hat{h}}_{jw} \\
		\mathbf{0} & \eta_2-P_{a}& -P_{a}\hat{h}_{aw} & \\
		-\zeta \mathbf{\hat{h}}_{jw}^H & -P_{a}\hat{h}_{aw} ^H & \eta_2\left ( -u_{jw}^2-\frac{v_{aw}^2}{M} \right )-c_2
	\end{bmatrix}\succeq \mathbf{0},
\end{equation}
where $\zeta =-\bar{b}P_{j,0}+P_{j,1}$ and $c_2 = \left (-\bar{b}P_{j,0}+P_{j,1}  \right )\left \|\mathbf{\hat{h}}_{jw}  \right \|^{2}+P_a\left| \hat{h}_{aw}\right| ^2-\left ( \bar{b}-1 \right )\sigma_{w}^{2}$.
\begin{equation}\label{eq:37}
	\begin{bmatrix}
		\left ( \eta_3+P_{j,1} \right )\mathbf{I}_M & \mathbf{0} & P_{j,1}\mathbf{\hat{h}}_{jw} \\
		\mathbf{0} & \eta_3-P_{a} & -P_{a}\hat{h}_{aw}\\
		P_{j,1}\mathbf{\hat{h}}_{jw}^H & -P_{a}\hat{h}_{aw}^H & \eta_3\left ( -u_{jw}^2-\frac{v_{aw}^2}{M} \right )-c_3
	\end{bmatrix}\succeq \mathbf{0},
\end{equation}
where $c_3=-P_{j,1}\left \|\mathbf{\hat{h}}_{jw}  \right \|^{2}+P_a\left| \hat{h}_{aw}\right| ^2$.\par
Therefore, we obtain the following conservative approximation of \eqref{eq:25} as
\begin{equation}\label{eq:38}
	\begin{split}
		&\underset{P_a,P_{j,0},P_{j,1}}{\mathrm{max}}P_a\\
		&s.t.\quad \eqref{eq:25d},\eqref{eq:25e},\eqref{eq:35},\eqref{eq:36},\eqref{eq:37}.	
	\end{split}
\end{equation}\par
It is clear that \eqref{eq:38} is a typical LMI optimal problem. Both the objective function and constraint functions are convex. Thus, the optimal solution for power allocation exists. Unlike Scenario 1, the analytical solution to the problem cannot be obtained. Using the optimization toolbox in MATLAB$^{\copyright }$, we can compute the numerical solutions of $P_a$, $P_{j,0}$ and $P_{j,1}$, and the maximum covert rate $R_b$ can be obtained.
%\emph{Disscussion:} Since we consider two channel estimation errors, it is necessary to find out which one affects the performance more. When $\sigma_{aw}^2$ increases, this leads to an increase in $\hat{h}_{aw}$. In this case, the effect of $\Delta h_{aw}$ is smaller than that of $\Delta \mathbf{h}_{jw}$. However, when Jammer is closer to Willie, the effect of $\Delta \mathbf{h}_{jw}$ is smaller than that of $\Delta h_{aw}$. Without considering the channel state, when $M$ increases, due to the boundary constraints model, both the impact of $\Delta \mathbf{h}_{jw}$ and $\Delta h_{aw}$ become smaller. Furthermore, from  \eqref{eq:27}, it is clear that the bound on $\left| \Delta h_{aw}\right| ^2$ also drops. After comparison, we find that the channel estimation error of $h_{aw}$ has less impact on the performance.
\subsection{Statistical CSI Scenario}
In high-security scenarios, such as military communications, Willie may be an enemy device and has an adversarial relationship with Alice, Jammer, and Bob. In this case, it is impractical to assume that Alice knows Willie's instantaneous CSI or even imperfect CSI. Therefore, we discuss the situation where only the statistical CSI, e.g., $\sigma _{aw}^{2}$ and $\sigma _{jw}^{2}$, are known. Mathematically, according to \eqref{eq:20}, the optimization problem is formulated as
\begin{subequations}\label{eq:39}
	\begin{align}
		&\underset{P_a,P_{j,0},P_{j,1}}{\mathrm{max}}P_a, \label{eq:39a}\\
		s.t. 	\  &\mathbb{E}\left\lbrace D\left ( \mathbb{P}_{0}\|\mathbb{P}_{1} \right  )\right\rbrace \leq \frac{2\varepsilon ^2}{L}, \label{eq:39b}\\
		&P_a\sigma_{aw}^2\leq P_{j,1}M\sigma_{jw}^2, \label{eq:39c}\\
		&P_{j,1}+P_{a} \leq P_{\mathrm{max}},\label{eq:39d}\\
		&P_{j,0} \leq P_{\mathrm{max}},\label{eq:39e}
	\end{align}
\end{subequations}
where $\mathbb{E}\left\lbrace .\right\rbrace $ denotes the statistical expectation.
Substituting $\lambda _0$ and $\lambda _1$ into $D\left ( \mathbb{P}_{0}\|\mathbb{P}_{1} \right  )$, we get
\begin{equation}\label{eq:40}
	\begin{split}
		&D\left ( \mathbb{P}_{0}\|\mathbb{P}_{1} \right  )\\
		=&\mathrm{ln}\frac{\left | h_{aw} \right |^2P_{a}+\left \|\mathbf{h}_{jw}  \right \|^2P_{j,1}+\sigma _{w}^{2}}{\left \|\mathbf{h}_{jw}  \right \|^2P_{j,0}+\sigma _{w}^{2}}\\
		&+\frac{\left \|\mathbf{h}_{jw}  \right \|^2P_{j,0}+\sigma _{w}^{2}}{\left | h_{aw} \right |^2P_{a}+\left \|\mathbf{h}_{jw}  \right \|^2P_{j,1}+\sigma _{w}^{2}}-1.
	\end{split}
\end{equation}
Since Alice and Jammer only know $\sigma _{aw}^{2}$ and $\sigma _{jw}^{2}$, the statistical expectation of $D\left ( \mathbb{P}_{0}\|\mathbb{P}_{1} \right  )$ is obtained by integrating over $\left | h_{aw} \right |^2$ and $\left \|\mathbf{h}_{jw}  \right \|^2$. Considering that each channel is assumed to follow the Rayleigh fading model, $\left | h_{aw} \right |^2$ follows an exponential distribution with a parameter of $\frac{1}{\sigma _{aw}^{2}}$. The probability density function (PDF) of $\left | h_{aw} \right |^2$ is
\begin{equation}\label{eq:41}
	f_{\left | h_{aw} \right |^2}\left ( x \right ) = \frac{1}{\sigma_{aw}^2}\exp\left ({-\frac{x}{\sigma_{aw}^2}}  \right ).
\end{equation}
Similarly, since $\left \|\mathbf{h}_{jw}  \right \|^2$ is the sum of $M$ independent exponential RV, it follows a Gamma distribution. The PDF of $\left \|\mathbf{h}_{jw}  \right \|^2$ is
\begin{equation}\label{eq:42}
	f_{\left \|\mathbf{h}_{jw}  \right \|^2}\left ( y \right ) = \frac{\left (\sigma_{jw}^2  \right )^{-M}}{\left ( M-1 \right )!}x^{M-1}\exp\left ({-\frac{y}{\sigma_{jw}^2}}  \right ).
\end{equation}
Using \eqref{eq:40}-\eqref{eq:42}, $\mathbb{E}\left\lbrace D\left ( \mathbb{P}_{0}\|\mathbb{P}_{1} \right  )\right\rbrace$ is calculated as
\begin{equation}\label{eq:43}
	\begin{split}
		&\mathbb{E}\left\lbrace D\left ( \mathbb{P}_{0}\|\mathbb{P}_{1} \right  )\right\rbrace \\
		=&\int_{0}^{\infty}\int_{0}^{\infty}D\left ( \mathbb{P}_{0}\|\mathbb{P}_{1} \right  )f_{\left | h_{aw} \right |^2}\left ( x \right )f_{\left \|\mathbf{h}_{jw}  \right \|^2}\left ( y \right ) dxdy\\
		=&\int_{0}^{\infty}\int_{0}^{\infty}\left [\mathrm{ln}\frac{xP_{a}+yP_{j,1}+\sigma _{w}^{2}}{yP_{j,0}+\sigma _{w}^{2}}+\frac{yP_{j,0}+\sigma _{w}^{2}}{xP_{a}+yP_{j,1}+\sigma _{w}^{2}}-1   \right ]\\
		&\times \left [\frac{1}{\sigma_{aw}^2}\exp\left ({-\frac{x}{\sigma_{aw}^2}}  \right )  \right ]\\
		&\times \left [\frac{\left (\sigma_{jw}^2  \right )^{-M}}{\left ( M-1 \right )!}x^{M-1}\exp\left ({-\frac{y}{\sigma_{jw}^2}}  \right )  \right ]dxdy.
	\end{split}
\end{equation}\par
Since there are logarithmic, exponential, and power functions in \eqref{eq:43}, this double integral is complicated and intractable. To solve this problem, we analyzed $D\left ( \mathbb{P}_{0}\|\mathbb{P}_{1} \right  )$ asymptotically and converted the double integral into a single integral by substitution, which significantly reduces the computational complexity and makes this problem solvable.\par
Asymptotic: For $D\left ( \mathbb{P}_{0}\|\mathbb{P}_{1} \right  )$, the denominator and numerator of both fractions in \eqref{eq:40} contain the noise power $\sigma _{w}^{2}$, which can be ignored when the noise power is considered to be very small compared to the jamming power. Since $\left | h_{aw} \right |^2>0$ and $ \left\| \mathbf{h}_{jw} \right\| ^2>0$, $D\left ( \mathbb{P}_{0}\|\mathbb{P}_{1} \right  )$ can be asymptotically expressed as
\begin{equation}\label{eq:44}
	\begin{split}
		D\left ( \mathbb{P}_{0}\|\mathbb{P}_{1} \right  )=&\mathrm{ln}\frac{xP_{a}+yP_{j,1}}{yP_{j,0}}+\frac{yP_{j,0}}{xP_{a}+yP_{j,1}}-1\\
		=&\mathrm{ln}\frac{\frac{x}{y}P_{a}+P_{j,1}}{P_{j,0}}+\frac{P_{j,0}}{\frac{x}{y}P_{a}+P_{j,1}}-1.\\
	\end{split}
\end{equation}
Let $\frac{x}{y}=z$, and the above formula is simplified as
\begin{equation}\label{eq:45}
	\begin{split}
		D\left ( \mathbb{P}_{0}\|\mathbb{P}_{1} \right  )=\mathrm{ln}\frac{zP_{a}+P_{j,1}}{P_{j,0}}+\frac{P_{j,0}}{zP_{a}+P_{j,1}}-1.
	\end{split}
\end{equation}\par
Double Integral Conversion: In the following, we employ a property of the F-distribution to obtain the PDF of RV $z$ , and convert (43) into a single integral operation.\\
\noindent {\bf{Property 1 (F-distribution)}}: Let $X \sim N\left ( \mu _1,\delta _1^2 \right )$ and $Y \sim N\left ( \mu _2,\delta _2^2 \right )$ be independent, Sample $\left ( X_1, X_2,..., X_{n_1}  \right )$ and sample $\left ( Y_1, Y_2,..., Y_{n_2}  \right )$ from $X$ and $Y$, respectively, we have
\begin{equation}\label{eq:46}
	\begin{split}
		F=\frac{\sum_{i=1}^{n_1}\left ( X_i-\mu_1 \right )^2}{n_1\delta_1^2}/\frac{\sum_{i=1}^{n_2}\left ( Y_i-\mu_2 \right )^2}{n_1\delta_2^2}\sim F\left ( n_1,n_2 \right ).
	\end{split}
\end{equation}
Since the exponential distribution and the Gamma distribution are both special chi-squared distributions, with the help of \textbf{Property 1}, we have
\begin{equation}\label{eq:47}
	\frac{2M\sigma_{jw}^2}{2\sigma_{aw}^2}z\sim F\left ( 2, 2M\right ),
\end{equation}
where the factors of $2$ and $2M$ are the degrees of freedom of $\left | h_{aw} \right |^2$ and $\left \|\mathbf{h}_{jw}  \right \|^2$, respectively. % The PDF of F-distribution is given by
%\begin{equation}\label{eq:48}
%	f\left ( x,n_1,n_2 \right )=\frac{\left ( n_1/n_2 \right )^{\frac{n_1}{2}}}{\mathrm{B}\left ( n_1/2, n_2/2 \right )}x^{\frac{n_1}{2}-1}\left ( 1+\frac{n_1}{n_2}x \right )^{-\frac{n_1+n_2}{2}}.
%\end{equation}
Thus, the PDF of $z$ can be written as
\begin{equation}\label{eq:49}
	\begin{split}	
		f_{Z}\left ( z \right )=&\frac{M\sigma _{jw}^{2}}{\sigma _{aw}^{2}}\frac{M^{-1}}{\mathrm{B}\left ( 1, M \right )}\left ( 1+\frac{1}{M}\frac{M\sigma _{jw}^{2}}{\sigma _{aw}^{2}}z \right )^{-\left ( M+1 \right )}\\
		=&M\left (\frac{\sigma _{aw}^{2}}{\sigma _{jw}^{2}}  \right )^M\left(\frac{\sigma _{aw}^{2}}{\sigma _{jw}^{2}}+z \right )^{-\left ( M+1 \right )}.
	\end{split}
\end{equation}
where $\mathrm{B}(x, y)$ is the beta function (Euler's integral of the first kind) and $\mathrm{B}(x, y)=\frac{\Gamma \left ( x \right )\Gamma \left ( y \right )}{\Gamma \left ( x+y \right )}$. Instead, with the help of \eqref{eq:45} and \eqref{eq:49}, let $\alpha = \frac{P_a}{P_{\max}}$, $\beta = \frac{P_{j,0}}{P_{\max}}$ and $\gamma = \frac{P_{j,1}}{P_{\max}}$, $\mathbb{E}\left\lbrace D\left ( \mathbb{P}_{0}\|\mathbb{P}_{1} \right  )\right\rbrace$ can be further expressed as
\begin{equation}\label{eq:50}
	\begin{split}
		&\mathbb{E}\left\lbrace D\left ( \mathbb{P}_{0}\|\mathbb{P}_{1} \right  )\right\rbrace \\
		=&\int_{0}^{\infty}D\left ( \mathbb{P}_{0}\|\mathbb{P}_{1} \right  )f_{Z}\left ( z \right )dz\\
		=&\int_{0}^{\infty}\left[ \mathrm{ln}\frac{\alpha z+\gamma }{\beta }+\frac{\beta}{\alpha z +\gamma}-1\right]\\
		&\times\left[ M\left (\frac{\sigma _{aw}^{2}}{\sigma _{jw}^{2}}  \right )^M\left(\frac{\sigma _{aw}^{2}}{\sigma _{jw}^{2}}+z \right)^{-\left ( M+1 \right )}\right] dz\\
		=&M\left (\frac{\sigma _{aw}^{2}}{\sigma _{jw}^{2}}  \right )^M\\
		&\times\int_{0}^{\infty}\left[ \mathrm{ln}\frac{\alpha z+\gamma }{\beta }+\frac{\beta}{\alpha z +\gamma}-1\right] \left(\frac{\sigma _{aw}^{2}}{\sigma _{jw}^{2}}+z \right)^{-\left ( M+1 \right )} dz.
	\end{split}
\end{equation}
The details of deriving $\mathbb{E}\left\lbrace D\left ( \mathbb{P}_{0}\|\mathbb{P}_{1} \right  )\right\rbrace$ are presented in Appendix A. With the help of Appendix A and $b=\frac{\sigma _{aw}^{2}}{\sigma _{jw}^{2}}$, we finally get the expression of $\mathbb{E}\left\lbrace D\left ( \mathbb{P}_{0}\|\mathbb{P}_{1} \right  )\right\rbrace$ as
\begin{equation}\label{eq:51}
	\begin{split}
		&\mathbb{E}\left\lbrace D\left ( \mathbb{P}_{0}\|\mathbb{P}_{1} \right  )\right\rbrace\\
		=&\mathrm{ln}\frac{\gamma }{\beta}+\frac{\alpha}{\gamma}b{M^{-1}}_2F_1\left ( 1,1;1+M;\frac{\gamma-\alpha b}{\gamma} \right )\\
		&	+\frac{\beta}{\gamma}{\frac{M}{M+1}}{_2F_1}\left ( 1,1;2+M;\frac{\gamma-\alpha b}{\gamma} \right )-1.
	\end{split}
\end{equation} \par
\emph{Discussion:} To further simplify the problem, the relationship between the transmission powers, especially between $P_{j,0}$ and $P_{j,1}$, is explored. Performing the partial derivative of \eqref{eq:51} with respect to $\gamma $, the result is give in \eqref{eq:52} at the bottom of next page. The detail of derivation is given in Appendix B. Setting $\frac{\partial }{\partial \gamma }\mathbb{E}\left\lbrace D\left ( \mathbb{P}_{0}\|\mathbb{P}_{1} \right  )\right\rbrace = 0$, we discover that when $M>1$, the second, fourth and fifth terms of the equation all tend to 0. Thus, $\gamma ^*\approx \beta $ is the root of $\frac{\partial }{\partial \gamma }\mathbb{E}\left\lbrace D\left ( \mathbb{P}_{0}\|\mathbb{P}_{1} \right  )\right\rbrace = 0$, regardless of $\alpha$. Moreover when $\gamma <\gamma ^*$,$\frac{\partial }{\partial \gamma }\mathbb{E}\left\lbrace D\left ( \mathbb{P}_{0}\|\mathbb{P}_{1} \right  )\right\rbrace < 0$, when $\gamma >\gamma ^*$,$\frac{\partial }{\partial \gamma }\mathbb{E}\left\lbrace D\left ( \mathbb{P}_{0}\|\mathbb{P}_{1} \right  )\right\rbrace > 0$. Thus, $\gamma ^*$  is the minimum point of the function $\mathbb{E}\left\lbrace D\left ( \mathbb{P}_{0}\|\mathbb{P}_{1} \right  )\right\rbrace $. Therefore the optimization problem has an optimal solution at $P_{j,0}=P_{j,1}$. Finally, considering $P_a+P_{j,1} \leq P_{\max}$, we get  $P_a=P_{\max}-P_{j,1}=P_{\max}-P_{j,0}$. After that, \eqref{eq:51} is asymptotically simplified as
\begin{figure*}[hb]
	\hrulefill
	\vspace*{10pt}\\
	\begin{equation}\label{eq:52}
		\begin{split}
			\frac{\partial }{\partial \gamma }\mathbb{E}\left\lbrace D\left ( \mathbb{P}_{0}\|\mathbb{P}_{1} \right  )\right\rbrace=&\frac{1}{\gamma }-\frac{\alpha b}{\gamma ^2}M^{-1}{_2F_1}\left ( 1,1;1+M;\frac{\gamma-\alpha b}{\gamma} \right )-\frac{\beta }{\gamma ^2}\frac{M}{M+1}{_2F_1}\left ( 1,1;2+M;\frac{\gamma-\alpha b}{\gamma} \right )\\
			&+\frac{\alpha ^2b^2}{\gamma ^3}\left [ M\left ( M+1 \right ) \right ]^{-1}{_2F_1}\left ( 2,2;2+M;\frac{\gamma-\alpha b}{\gamma} \right )+\frac{\alpha \beta b}{\gamma ^3}\frac{M}{\left (M+1  \right )^2}{_2F_1}\left ( 2,2;3+M;\frac{\gamma-\alpha b}{\gamma} \right )	
		\end{split}
	\end{equation}
\end{figure*}\par

\setcounter{equation}{51}
\begin{equation}\label{eq:53}
	\begin{split}
		\mathbb{E}\left\lbrace D\left ( \mathbb{P}_{0}\|\mathbb{P}_{1} \right  )\right\rbrace=&\frac{\alpha}{\gamma}b{M^{-1}}_2F_1\left ( 1,1;1+M;\frac{\gamma-\alpha b}{\gamma} \right )\\
		&+{\frac{M}{M+1}}{_2F_1}\left ( 1,1;2+M;\frac{\gamma-\alpha b}{\gamma} \right )-1,
	\end{split}
\end{equation}
where $\alpha=1-\gamma$. {Meanwhile, by taking the second partial derivative of $\alpha $ in \eqref{eq:53}, it is proved that \eqref{eq:53} is a convex function. Therefore, \eqref{eq:39} is converted from a multivariable optimization problem to a univariate optimization problem which can be solved by CVX in Matlab$^{\copyright }$. Observing \eqref{eq:53}, it is clear that for the same statistical CSI, the power allocated to Alice is increased with $M$.}\par

\section{Special Case}
As the performance benchmark of the proposed cooperative deception strategy, we study the scenario where Jammer does not participate in cooperative deception. In this case, Willie will use MRC on Alice under $\mathcal{H}_1$. Since Jammer has no deceptive effect under $\mathcal{H}_1$, and $P_{j,1}> 0$ will consume Alice's transmit power and cause interference to Bob's receptions. Therefore, we assume that in this case, Jammer only transmits AN under $H_0$, and Alice only transmits covert information at  $H_1$. Under this assumption, the covert transmission performance under three different CSI scenarios is studied, and the covert rate is also maximized through power allocation.
\subsection{Instantaneous CSI Scenario}
In this scenario, similar to \eqref{eq:18}, the optimization problem is expressed as follows
\begin{equation}\label{eq:54}
	\begin{split}
		\underset{P_a,P_{j,0}}{\mathrm{max}}&{R_{b}}^{c},\\
		s.t. 	\ &(18b),\\
				& P_a,P_{j,0} \leq P_{\max},
	\end{split}
\end{equation}
Since the transmission strategy has changed, the expression of $D\left ( \mathbb{P}_{0}\|\mathbb{P}_{1} \right)$ is also changed as
\begin{equation}\label{eq:55}
	D\left ( \mathbb{P}_{0}\|\mathbb{P}_{1} \right ) =\ln\frac{P_{a}\left \|\mathbf{h}_{aw}  \right \|^2+\sigma _{w}^{2}}{P_{j,0}\left \|\mathbf{h}_{jw}  \right \|^2+\sigma _{w}^{2}}+\frac{P_{j,0}\left \|\mathbf{h}_{jw}  \right \|^2+\sigma _{w}^{2}}{P_{a}\left \|\mathbf{h}_{aw}  \right \|^2+\sigma _{w}^{2}}-1.
\end{equation}
Obviously, \eqref{eq:18b} can be transformed into the form
\begin{equation}\label{eq:56}
	\begin{split}
		P_a\left \|\mathbf{h}_{aw}  \right \|^2 &= P_{j,0}\left \|\mathbf{h}_{jw}  \right \|^2\\
		\mathrm{i.e.}, P_a\left| h_{aw}\right|^2 &= P_{j,0}\left| h_{jw}\right|^2.
	\end{split}
\end{equation}
\emph{Discussion: }It can be seen from \eqref{eq:56} that the covert transmission performance in this case has nothing to do with the number of Willie's antennas $M$, but is only related to the transmission power as well as the channel parameters $\left | h_{aw} \right |^2$ and $\left | h_{jw} \right |^2$.
\subsubsection{Case of $\left | h_{aw} \right |^2\geq \left | h_{jw} \right |^2$}
To maximize Alice's transmit power, $P_{j,0}$ should be $P_{\max}$, so that we get
\begin{equation}\label{eq:57}
	P_{a}^* = \frac{P_{\mathrm{max}}\left | h_{jw} \right |^2}{\left | h_{aw} \right |^2}\ \mathrm{and}\ P_{j,0}^* = P_{\max}.
\end{equation}
\subsubsection{Case of $\left | h_{aw} \right |^2 < \left | h_{jw} \right |^2$}
Since $P_{j,0}$ participates in optimization, in this case, $P_a$ can reach $P_{\max}$ by adjusting $P_{j,0}$, which means that
\begin{equation}\label{eq:58}
	P_{a}^* = P_{\max}\ \mathrm{and}\ P_{j,0}^* = \frac{P_{\mathrm{max}}\left | h_{aw} \right |^2}{\left | h_{jw} \right |^2}.
\end{equation}
{From \eqref{eq:58}, since $P_{j,1}=0$ in the benchmark and $P_a = P_{\max}$, the covert rate can achieve its maximum value, $R_{\max}=\mathrm{log}\left ( 1+\frac{P_{\max} \left | h_{ab} \right |^2}{\sigma _{b}^{2}} \right )$.}
\subsection{Imperfect CSI Scenario}
Under imperfect CSI, the optimization problem is formulated as
\begin{equation}\label{eq:59}
	\begin{split}
		&\underset{P_a,P_{j,0}}{\mathrm{max}}P_a\\
		&s.t.\quad (25b),(25f),(25g),
	\end{split}
\end{equation}
where \eqref{eq:25b} can be equivalently written as
\begin{subequations}\label{eq:60}
	\begin{align}
		\bar{a}P_{j,0}\left \|\mathbf{h}_{jw}  \right \|^2 -P_{a}\left \|\mathbf{h}_{aw}  \right \|^2+\left ( \bar{a}-1 \right )\sigma _{w}^{2}&\leq 0,\\
		-\bar{b}P_{j,0}\left \|\mathbf{h}_{jw}  \right \|^2 +P_{a}\left \|\mathbf{h}_{aw}  \right \|^2-\left ( \bar{b}-1 \right )\sigma _{w}^{2}&\leq 0.
	\end{align}
\end{subequations}
Similar to the solution of cooperative deception, Equations \eqref{eq:25b}, \eqref{eq:25f} and \eqref{eq:25g} can be rewritten in matrix form as
\begin{subequations}\label{eq:61}
	\begin{align}
	\begin{bmatrix}
		-P_a\mathbf{I}_{M} & 0\cdot\mathbf{I}_{M}  & -P_a\mathbf{\hat{h}}_{aw}\\
		0\cdot\mathbf{I}_{M} & \bar{a}P_{j,0}\mathbf{I}_{M}  & \bar{a}P_{j,0}\mathbf{\hat{h}}_{jw}\\
		-P_a\mathbf{\hat{h}}_{aw}^H & \bar{a}P_{j,0}\mathbf{\hat{h}}_{jw}^H & d_1 \\
	\end{bmatrix}
	&\preceq  \textbf{0},\label{eq:61a}\\
	\begin{bmatrix}
		P_a\mathbf{I}_{M} & 0\cdot\mathbf{I}_{M}  & P_a\mathbf{\hat{h}}_{aw}\\
		0\cdot\mathbf{I}_{M} & -\bar{b}P_{j,0}\mathbf{I}_{M}  & -\bar{b}P_{j,0}\mathbf{\hat{h}}_{jw}\\
	P_a\mathbf{\hat{h}}_{aw}^H & -\bar{b}P_{j,0}\mathbf{\hat{h}}_{jw}^H & d_2 \\
	\end{bmatrix}
	&\preceq  \textbf{0}\ \mathrm{and} \label{eq:61b}\\
	\begin{bmatrix}
		\mathbf{I}_{M} & 0\cdot\mathbf{I}_{M}  & \mathbf{0}\\
		0\cdot\mathbf{I}_{M} & \mathbf{I}_{M} & \mathbf{0}\\
		\mathbf{0} & \mathbf{0} & -v_{aw}^2-u_{jw}^2
	\end{bmatrix}
	&\preceq  \textbf{0}\label{eq:61c}
	\end{align}
\end{subequations}
where $d_1=-P_a\left \|\mathbf{\hat{h}}_{aw}  \right \|^{2}+\bar{a}P_{j,0}\left \|\mathbf{\hat{h}}_{jw}  \right \|^{2}+\left ( \bar{a}-1 \right )\sigma_{w}^{2}$ and $d_2=P_a\left \|\mathbf{\hat{h}}_{aw}  \right \|^{2}-\bar{b}P_{j,0}\left \|\mathbf{\hat{h}}_{jw}  \right \|^{2}-\left ( \bar{b}-1 \right )\sigma_{w}^{2}$.
Use the deformation of the S-Procedure to deal with \eqref{eq:61a}, \eqref{eq:61b} and \eqref{eq:61c}. The implications \eqref{eq:61c}$\Rightarrow $\eqref{eq:61a} and \eqref{eq:61c}$\Rightarrow $\eqref{eq:61b} hold if and only if there exist variables $\phi_1 >0$ and $\phi_2 >0$ such that
\begin{equation}\label{eq:62}
	\begin{bmatrix}
		\left (\phi_1 +P_a  \right )\mathbf{I}_{M} & 0\cdot\mathbf{I}_{M}  & P_a\mathbf{\hat{h}}_{aw}\\
		0\cdot\mathbf{I}_{M} & \left (\phi_1 -\bar{a}P_{j,0}  \right )\mathbf{I}_{M} & -\bar{a}P_{j,0}\mathbf{\hat{h}}_{jw}\\
		P_a\mathbf{\hat{h}}_{aw}^H & -\bar{a}P_{j,0}\mathbf{\hat{h}}_{jw}^H & \phi_1\varphi -d_1
	\end{bmatrix}\succeq \textbf{0}
\end{equation}
and
\begin{equation}\label{eq:63}
	\begin{bmatrix}
		\left (\phi_2 -P_a  \right )\mathbf{I}_{M} & 0\cdot\mathbf{I}_{M}  & -P_a\mathbf{\hat{h}}_{aw}\\
		0\cdot\mathbf{I}_{M} & \left (\phi_2 +\bar{b}P_{j,0}  \right )\mathbf{I}_{M} & \bar{b}P_{j,0}\mathbf{\hat{h}}_{jw}\\
		P_a\mathbf{\hat{h}}_{aw}^H & -\bar{b}P_{j,0}\mathbf{\hat{h}}_{jw}^H & \phi_2\varphi -\iota
	\end{bmatrix}\succeq \textbf{0}
\end{equation}
where $\varphi = -v_{aw}^2-u_{jw}^2$.
Finally, \eqref{eq:59} can be written as
\begin{equation}\label{eq:64}
	\begin{split}
		&\underset{P_a,P_{j,0}}{\mathrm{max}}P_a\\
		&s.t.\quad \eqref{eq:62}, \eqref{eq:63}.			
	\end{split}
\end{equation}
Like \eqref{eq:38}, \eqref{eq:64} also can be solved using the optimization toolbox in Matlab$^{\copyright }$. \par
%\emph{Discussion:} From the expression of $\varphi $, it is clear that swap $u_{aw}^2$ and $v_{jw}^2$ does not change the constraints of \eqref{eq:64}, which means $\Delta\mathbf{h}_{aw}$ and $\Delta\mathbf{h}_{jw}$ have the same degree of influence on the optimization problem.
%From \eqref{eq:61a}, since Willie use MRC both on Jammer and Alice, it is clear that when $M$ increases, $\Delta\mathbf{h}_{aw}$ and $\Delta\mathbf{h}_{jw}$ will be affected similarly. {How effect??????????????????????????????????????????detail}.
\subsection{Statistical CSI Scenario}
For this scenario, the optimization problem is expressed as
\begin{equation}\label{eq:65}
	\begin{split}
		&\underset{P_a,P_{j,0}}{\mathrm{max}}P_a, \\
		s.t. 	\  &\mathbb{E}\left\lbrace D\left ( \mathbb{P}_{0}\|\mathbb{P}_{1} \right  )\right\rbrace \leq \frac{2\varepsilon ^2}{L}, \\
		&P_{j,0},P_a \leq P_{\mathrm{max}}.
	\end{split}
\end{equation}
Although under this scenario, $P_{j,1} = 0$, due to the existence of $\left \|\mathbf{h}_{aw}  \right \|^2$ and $\left \|\mathbf{h}_{jw}  \right \|^2$, we also have the problem of double integration. Therefore, refer to the method in Section \uppercase\expandafter{\romannumeral3}.\textit{C} and ignore $\sigma_w^2$.  Moreover, let $\left \|\mathbf{h}_{aw}  \right \|^2=x$, $\left \|\mathbf{h}_{jw}  \right \|^2 = y$, $\frac{x}{y} = z'$ and $\frac{P_a}{P_{j,0}}=\alpha$, then the covert constraint $D\left ( \mathbb{P}_{0}\|\mathbb{P}_{1} \right  )$ can be expressed as
\begin{equation}\label{eq:66}
	\begin{split}
		D\left ( \mathbb{P}_{0}\|\mathbb{P}_{1} \right  )=&\mathrm{ln}\frac{xP_{a}}{yP_{j,0}}+\frac{yP_{j,0}}{xP_{a}}-1\\
		=&\mathrm{ln}\left( \alpha z'\right) +\left( \alpha z'\right)^{-1}-1.\\
	\end{split}
\end{equation}
Similar to \eqref{eq:50}, with the help of \eqref{eq:49}, we have
\begin{equation}\label{eq:67}
	\frac{\sigma_{jw}^2}{\sigma_{aw}^2}z'\sim F\left ( 2M, 2M\right ),
\end{equation}
where $2M$ is the degree of freedom of $\left \|\mathbf{h}_{aw}  \right \|^2$ and $\left \|\mathbf{h}_{jw}  \right \|^2$. The PDF of $z'$ can be written as
\begin{equation}\label{eq:68}
	\begin{split}	
		f_{Z^{'}}\left ( z' \right )=&\frac{\sigma _{jw}^{2}}{\sigma _{aw}^{2}}\frac{1}{\mathrm{B}\left ( M, M \right )}\\
		&\times \left (\frac{\sigma _{jw}^{2}}{\sigma _{aw}^{2}}z'  \right )^{M-1}\left ( 1+\frac{\sigma _{jw}^{2}}{\sigma _{aw}^{2}}z' \right )^{-2M}\\
		=&\frac{1}{\mathrm{B}\left ( M, M \right )}\\
		&\times \left (\frac{\sigma _{aw}^{2}}{\sigma _{jw}^{2}}  \right )^M\left (z'  \right )^{M-1}\left ( \frac{\sigma _{aw}^{2}}{\sigma _{jw}^{2}} +z' \right )^{-2M}.
	\end{split}
\end{equation}
Let $b' = \frac{\sigma _{aw}^{2}}{\sigma _{jw}^{2}}$, then $\mathbb{E}\left\lbrace D\left ( \mathbb{P}_{0}\|\mathbb{P}_{1} \right  )\right\rbrace$ can be further expressed as
\begin{equation}\label{eq:69}
	\begin{split}
		&\mathbb{E}\left\lbrace D\left ( \mathbb{P}_{0}\|\mathbb{P}_{1} \right  )\right\rbrace \\
		=&\int_{0}^{\infty}D\left ( \mathbb{P}_{0}\|\mathbb{P}_{1} \right  )f_{Z^{'}}\left ( z' \right )dz'\\
		=&\frac{{b'}^M}{\mathrm{B}\left ( M, M \right )}\int_{0}^{\infty}D\left ( \mathbb{P}_{0}\|\mathbb{P}_{1} \right  )\left (z'  \right )^{M-1}\left ( b' +z' \right )^{-2M}dz'\\
		=&\frac{{b'}^M}{\mathrm{B}\left ( M, M \right )}\left( A'+B'+C'\right).
	\end{split}
\end{equation}
With the help of the following equation \cite{2007247}
%[31] »ý·Ö±í FI II 775a, ET I 310(19)
\begin{equation}\label{eq:70}
	\begin{split}
	\int_{0}^{\infty}\frac{x^{\mu -1}dx}{\left ( 1+\beta x \right )^\nu }=&\beta ^{-\mu }\mathrm{B}\left ( \mu, \nu -\mu \right ),\\
	&\left [ \left | \mathrm{arg}\beta \right |<\pi,  \mathrm{Re}\ \nu  >\mathrm{Re}\ \mu> 0 \right ]
	\end{split}
\end{equation}
we get
\begin{equation}\label{eq:71}
%	\begin{split}
		A' =\int_{0}^{\infty}\mathrm{ln}{\left( \alpha z'\right) }\left (z'  \right )^{M-1}\left ( b' +z' \right )^{-2M}dz',
%	\end{split}
\end{equation}
\begin{equation}\label{eq:72}
	\begin{split}
		B' =&\int_{0}^{\infty}\alpha ^{-1}{z'}^{M-2}\left(b'+z' \right)^{-2M}dz'\\
		   =&\frac{\mathrm{B}\left(M-1, M+1 \right) }{\alpha {b'}^{M+1}}\  \mathrm{and}
	\end{split}
\end{equation}
\begin{equation}\label{eq:73}
	\begin{split}
			C' =&-\int_{0}^{\infty}{z'}^{M-1}\left(b'+z' \right)^{-2M}dz'\\
			   =&-\frac{\mathrm{B}\left ( M, M \right )}{{b'}^M}.
	\end{split}
\end{equation}\par
Although $\mathbb{E}\left\lbrace D\left ( \mathbb{P}_{0}\|\mathbb{P}_{1} \right  )\right\rbrace$ is a multivariate optimization problem, the only constraint with variable coupling is only related to $\alpha = \frac{P_a}{P_{j,0}}$. Thus, \eqref{eq:65} can be transformed into a univariate $\alpha$ optimization problem. The convexity of the constraints can be proved using the second-order derivative, and the optimal solution can be obtained using CVX.

\section{Numerical result}
In this section, we present and discuss the numerical performance results for our proposed transmission strategy. Monte-Carlo simulations are performed using 10,000 samples. To gain deep insight into our work, we assume that Jammer is not participating in deception, e.g., $P_{j,1}=0$, as a benchmark, which further illustrates the reliability of proposed cooperative deception strategy. In our simulations, we set the number of symbols in a time slot $L = 50$, $P_{\max}=2$W, $\sigma_{ab}^2=\sigma_{jb}^2 =1$ and $\sigma_{b}^2 = \sigma_{w}^2= 10^{-3}$. In the following, we present some numerical examples of how different parameters,  $\left\| \Delta \mathbf{h}_{aw} \right\|^2$, $\left\| \Delta \mathbf{h}_{jw} \right\|^2$ and $M$, affect the covert transmission performance.

\subsection{Discussion for Scenario 1}
\begin{figure}[!t]
	\centering
	\includegraphics[width=3.2in]{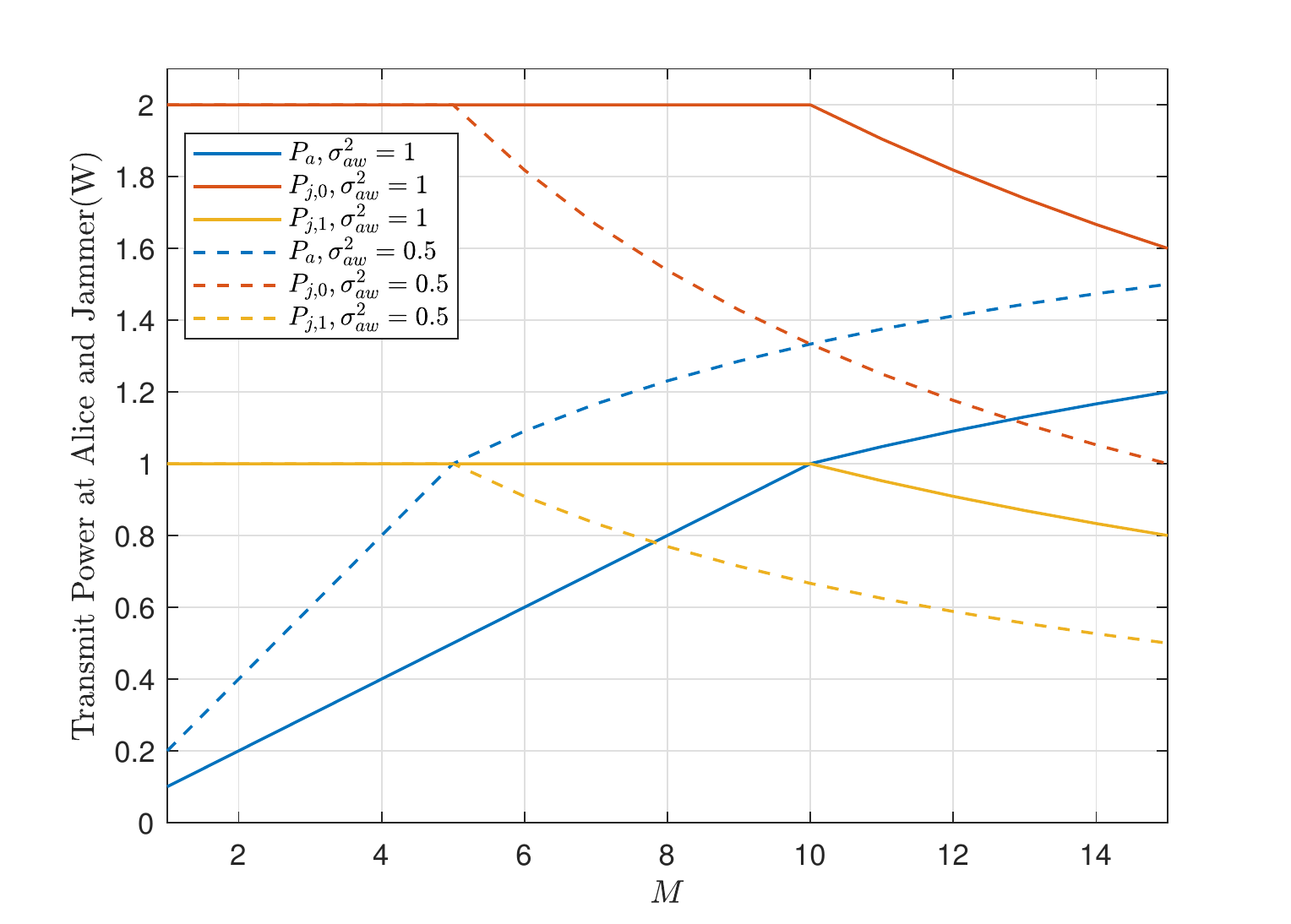}
	\caption{The transmit power versus $M$ with different channel gains under instantaneous CSI ($\sigma_{aw}^2>\sigma_{jw}^2$), $\sigma_{jw}^2=0.1$.}
	\label{fig_2}
\end{figure}

In this case, Alice knows the instantaneous CSI and Alice's transmission can be completely covert. Fig. \ref{fig_2} illustrates the impact of $M$ on the transmission power at Alice and Jammer when $\sigma_{aw}^2>\sigma_{jw}^2$. It can be found when $M$ is small, $P_a$ increases linearly with $M$ while $P_{j,0}$ and $P_{j,1}$ remain unchanged. This is because Willie uses MRC for Jammer, and Willie's detection of Jammer becomes stronger with the increase of $M$,  which is more helpful for Alice's transmission. However, when $M$ is large, $P_a$ increases slightly while $P_{j,0}$ and $P_{j,1}$ drop gradually. {This behavior can be explained using \eqref{eq:20b} and \eqref{eq:20d}, where the received power for ${\mathcal{H}_0}$ is required to be equal to that for ${\mathcal{H}_0}$ due to the perfect covert constraint. Also, the simulation matches our analysis in the \emph{Discussion} of Section \uppercase\expandafter{\romannumeral3}.A.} In addition, for the same $M$, we find that the decrease of $\sigma_{aw}^2$ weakens Willie's detection of Alice's transmission, which is the same as traditional recognition. As a result, $P_a$ increases and the covert rate is improved.\par
%\begin{figure}[!t]
%	\centering
%	\subfloat[]{\includegraphics[width=3.0in]{1_perfect_CSI_power_allocation_update.pdf}%
%		\label{fig_first_case}}
	%???$D\left ( p_{0}\parallel  p_{1}\right )=0$, $P_{j,0} = P_{\max}, \sigma_{jw}^2=0.1$ aw>jw
%	\hfil
%	\\
%	\subfloat[]{\includegraphics[width=3.0in]{2_perfect_CSI_power_allocation_update.pdf}%
%		\label{fig_second_case}}
	%???$D\left ( p_{0}\parallel  p_{1}\right )=0$, $ \sigma_{jw}^2=1$
%	\caption{Alice's and Jammer's transmit power versus the number of antennas of Willie under different channel gains with instantaneous perfect CSI. (a) $\sigma_{aw}^2>\sigma_{jw}^2$. (b) $\sigma_{aw}^2<\sigma_{jw}^2$.}
%	\label{fig_sim}
%\end{figure}

\begin{figure}[!t]
	\centering
	\includegraphics[width=3.2in]{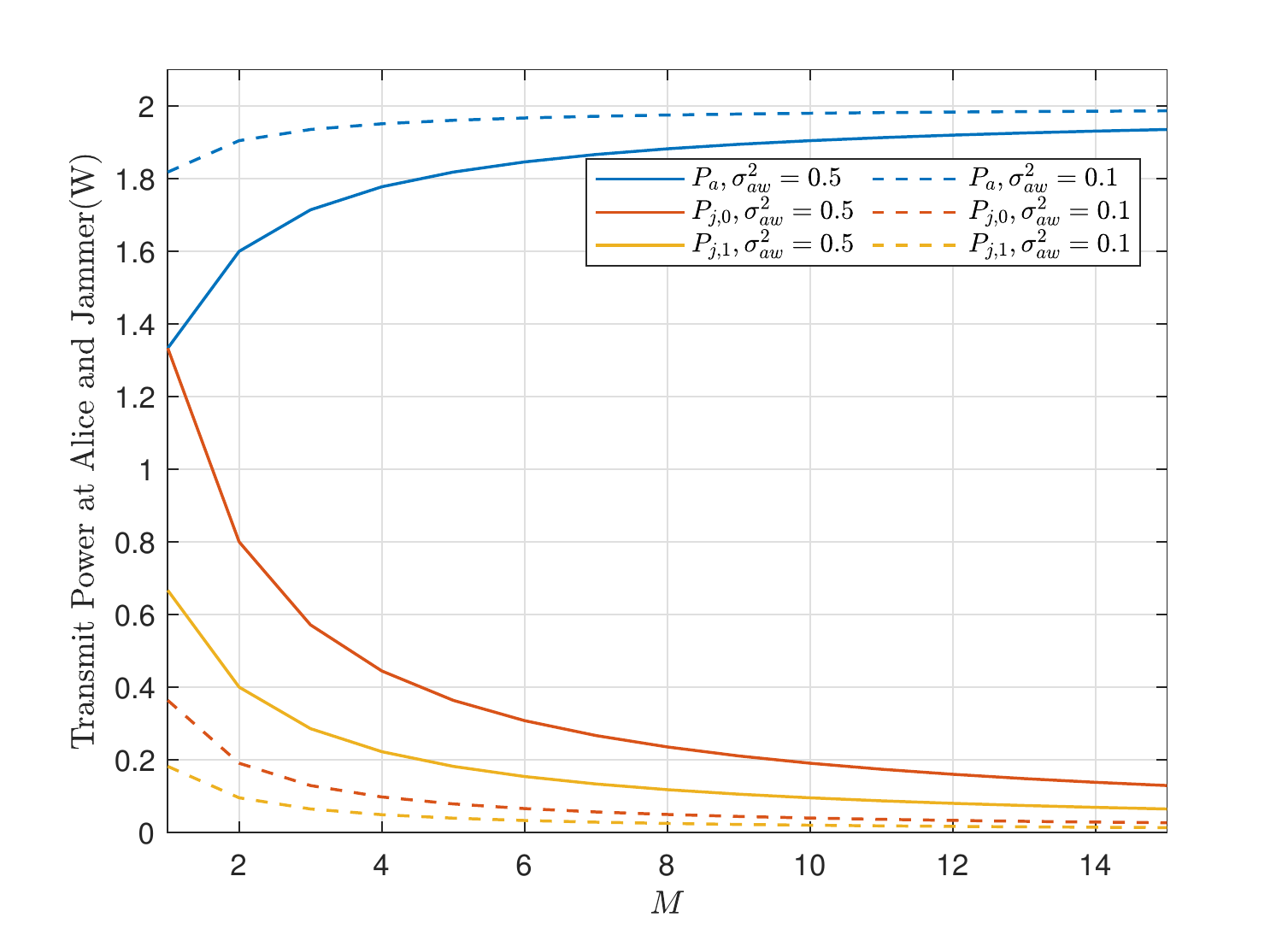}
	\caption{The transmit power versus $M$ with different channel gains under instantaneous CSI ($\sigma_{aw}^2<\sigma_{jw}^2$), $\sigma_{jw}^2=1$.}
	\label{fig_3}
\end{figure}
Figure \ref{fig_3} shows the impact of $M$ on the transmission power at Alice and Jammer when $\sigma_{aw}^2<\sigma_{jw}^2$. When $M$ increases, Jammer's transmit power decreases, and Alice's transmission power increases gradually. In this case, Willie has a strong detection for Jammer while a weak detection for Alice. Comparing Fig.\ref{fig_2} with Fig. \ref{fig_3}, we find that the transmission power at Alice in the case of $\sigma_{aw}^2<\sigma_{jw}^2$ is larger than that of $\sigma_{aw}^2>\sigma_{jw}^2$. This is due to the fact that in the case of $\sigma_{aw}^2<\sigma_{jw}^2$,  less power is used to deceive Willie and more power can be allocated to Alice for covert transmission.
With the help of Jammer's deception, the covert transmission performance is enhanced. Moreover, when the value of $\sigma_{aw}^2$ is small, Willie's detection for Alice is weaker and Alice's transmission power can be increased. \par
\begin{figure}[!t]
	\centering
	\includegraphics[width=3.2in]{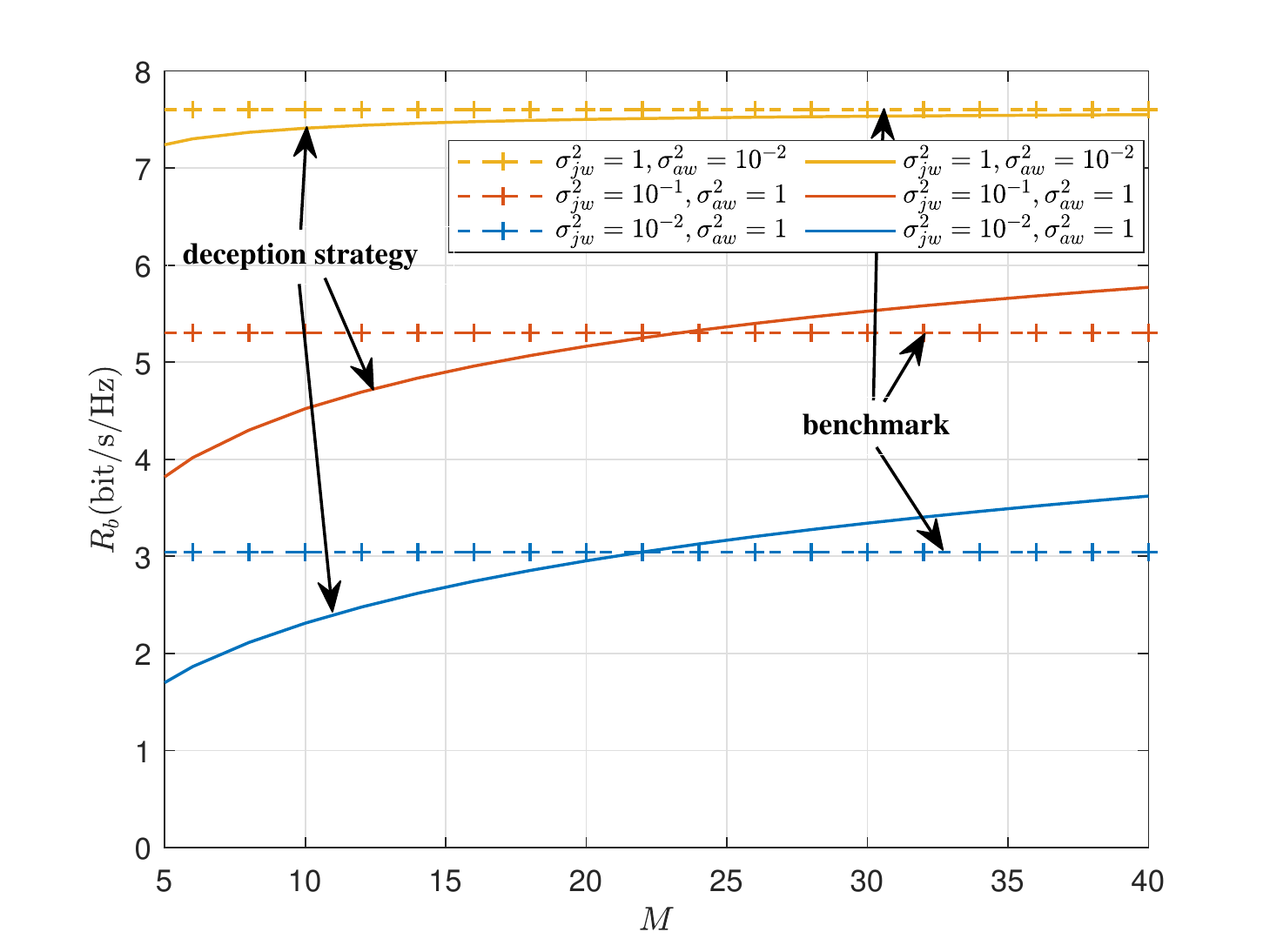}
	\caption{The covert rate versus $M$ under instantaneous CSI.}
	\label{fig_4}
\end{figure}
In Fig. \ref{fig_4}, we investigate the covert rates when the proposed deception strategy is applied or not. We find that the covert performance of the deception strategy is not necessarily better than that of the benchmark. When $\sigma_{aw}^2>\sigma_{jw}^2$ and $M$ is small, the covert rate of the deception strategy is less than that of the benchmark. Although the deception strategy weakens Willie's detection of Alice, it also introduces the noise to Bob's reception. Moreover, to deceive Willie successfully, we assume that in the proposed deception scheme the signal from Jammer should be larger than that from Alice. Thus, due to $M$ being small and $\sigma_{aw}^2>\sigma_{jw}^2$, there is no significant deception effect to assist Alice's transmission. However, when $\sigma_{aw}^2>\sigma_{jw}^2$ and $M$ is large, the covert transmission rate of the deception strategy outperforms that of the benchmark. Due to a large number of antennas, Jammer's deception works greatly, which allows Alice to transmit with larger power. In addition, when $\sigma_{aw}^2<\sigma_{jw}^2$, the covert performance of the deception strategy gradually approaches that of the benchmark as $M$ increases. Willie's detection of Alice under $\mathcal{H}_1$ is weaker than that of Jammer under $\mathcal{H}_0$. As a result, for the benchmark, even though Alice transmits with high power and MRC is used to receive Alice's signal, Alice's power received at Willie in $\mathcal{H}_1$ is less than Jammer's power received at Willie in $\mathcal{H}_0$. Moreover, without the noise caused by Jammer, the covert rate of the benchmark reaches its upper bound. {Hence, when $\sigma_{aw}^2<\sigma_{jw}^2$, the deception strategy cannot work better than the benchmark which matches our analysis in the \emph{Discussion} of Section \uppercase\expandafter{\romannumeral4}. A.} Moreover, we conclude that for the instantaneous CSI case, the deception strategy outperforms the benchmark when $\sigma_{aw}^2>\sigma_{jw}^2$ and $M$ is large.

\subsection{Discussion for Scenario 2}
In this scenario, Alice's transmission can not be completely covert, instead, the constraint is $D\left ( \mathbb{P}_{0}\|\mathbb{P}_{1} \right  )\leq \frac{2\varepsilon^2}{L}$. In this subsection, we discuss how the channel estimation errors $\left\| \Delta \mathbf{h}_{aw} \right\|^2$, $\left\| \Delta \mathbf{h}_{jw} \right\|^2$ and  $M$ affect the performance.\par
\begin{figure}[!t]
	\centering
	\includegraphics[width=3.2in]{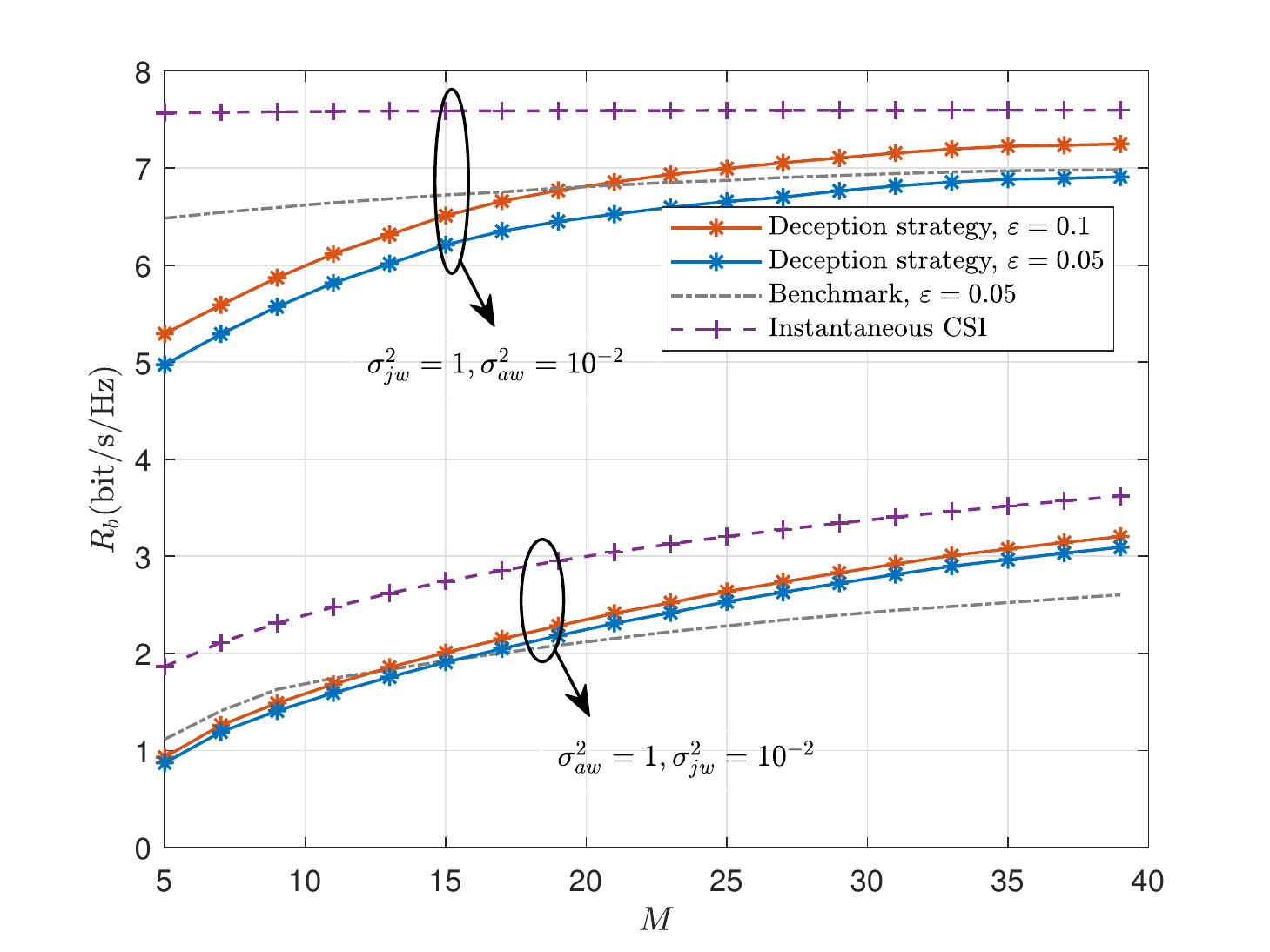}
	\caption{The covert rate versus $M$ with different channel gains under imperfect CSI, $v_{aw}^2 =u_{jw}^2 =10^{-4}$.}
	\label{fig_5}
\end{figure}
%\begin{figure}[!t]
%	\centering
%	\subfloat[]{\includegraphics[width=3.0in]{5_imperfect_CSI_R.pdf}%
%		\label{fig_first_case}}
%	\hfil
%	\\
	%Imperfect CSI $\sigma_{jb}^2 = 10^{-6},\sigma_{ab}^2 = 10^{-4}$,$\sigma_{w}^2 =\sigma_{b}^2  = 10^{-7}$,$v_{aw}^2 =u_{jw}^2 =10^{-8}$
%	\subfloat[]{\includegraphics[width=3.0in]{6_imperfect_CSI_R.pdf}%
%		\label{fig_second_case}}
%	\caption{$R_b$ versus Willie's number of antennas under different channel gains with imperfect CSI. (a) our proposed strategy (b) comparative case}
%	\label{fig_sim}
%\end{figure}

The impacts of $M$ and $\varepsilon $ on $R_b$ under imperfect CSI are shown in Figure \ref{fig_5}. {We find that when $M$ increases, the covert rate under imperfect CSI approaches that under instantaneous CSI. In this case, the estimation errors are bounded, though the estimated CSI $\mathbf{\hat{h}}_{jw}$ increases exponentially with $M$, which is explained in \eqref{eq:16} and \eqref{eq:17}. Therefore, the effects of channel estimation errors decrease gently with the increase of $M$. This behavior is also shown in the benchmark, as illustrated in Fig. \ref{fig_5}. The adverse effect of estimation errors decreases gently as $M$ increases for the same reason.} Moreover, when $\varepsilon $ increases, the covert transmission performance gets better. Because the relaxation of Willie's detection constraint allows Alice to transmit with a higher power. Additionally, for the same $\varepsilon $, comparing the benchmark to the deception strategy, we can arrive at the same conclusion as that under instantaneous CSI, which demonstrates that for the imperfect CSI case, the deception strategy is effective and can improve the covert rate when $\sigma_{aw}^2>\sigma_{jw}^2$ and $M$ is large.\par

%Due to channel estimation errors, it can be seen from Fig. 4 that, as expected, the covert rate with imperfect CSI is less than that with perfect CSI. However, by relaxing the constraints, i.e., as $\varepsilon$ increases, the covert rate with imperfect CSI is close to that with perfect CSI. {\color{blue}The estimation error is bounded, while the channel estimation parameter will increase exponentially as $M$ increases. Therefore, as the number of antennas $M$ increases, the influence of estimation error decreases. As a result, the covert rate  is increased and approaches the performance with perfect CSI. It can be seen from (b) that it is different from the case where the transmission capacity is independent of the number of antennas with Perfect CSI. The existence of channel estimation errors leads to the fact that the covert transmission performance with imperfect CSI is related to $M$. However, as in Scenario 1, as $M$ increases, the effect is gradually canceled out. The same as (a), with the same number of antennas, the performance improves as the covert constraints are relaxed. In addition, by comparing (a) and (b), we can obtain the same conclusion as in the perfect CSI scenario}.\par
\begin{figure}[!t]
	\centering
	\includegraphics[width=3.2in]{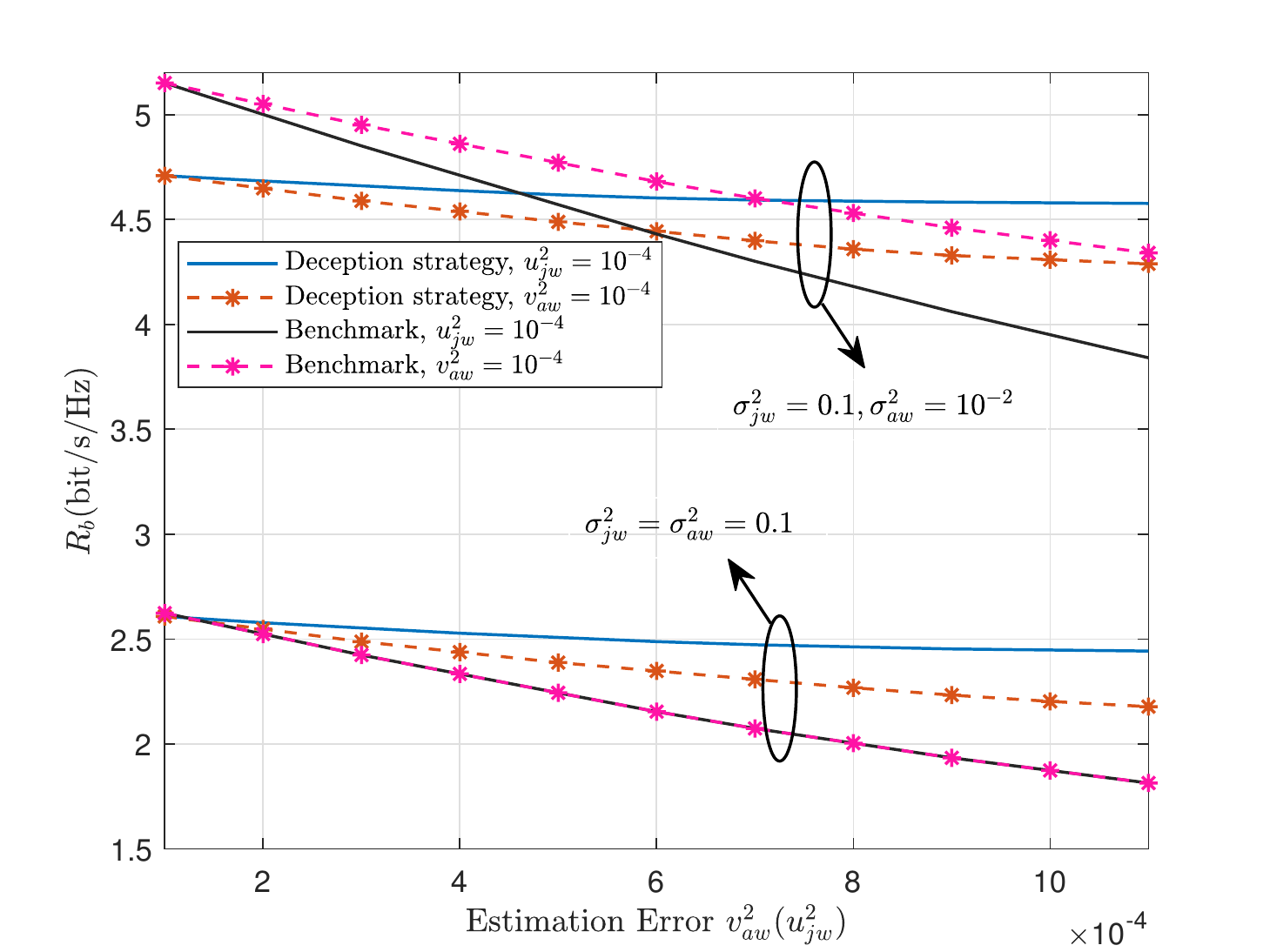}
	\caption{The performance of transmission versus the CSI estimation errors, $M$=20, $\varepsilon=0.05$}
	\label{fig_6}
\end{figure}

{Figure \ref{fig_6} investigates the covert rate against CSI estimation errors of both Alice-to-Willie and Jammer-to-Willie links. When $\sigma_{aw}^2=\sigma_{jw}^2$, we find that for the deception strategy, the influence of $u_{jw}^2$ is greater than that of $v_{aw}^2$. Willie uses MRC for Jammer instead of  Alice in this case, which results in only a single antenna gain at Willie for Alice's transmission. Therefore, the influence of $v_{aw}^2$ decreases to that of $\frac{v_{aw}^2}{M}$, which is explained by \eqref{eq:27}. However, for the benchmark, $u_{jw}^2$ and $v_{aw}^2$ have the same effect on the performance when $\sigma_{aw}^2=\sigma_{jw}^2$. In this case, Willie uses MRC for Jammer under $\mathcal{H}_0$ and for Alice under $\mathcal{H}_1$, respectively. In addition, the case of $\sigma_{aw}^2<\sigma_{jw}^2$ is also illustrated in Fig. \ref{fig_6}. We can find that, for the deception strategy, even if $\sigma_{aw}^2$ decreases, the influence of $v_{aw}^2$ is still less than that of $u_{jw}^2$ for $M\gg 1$. However, for the benchmark, without the help of Jammer's deception, the influence of $v_{aw}^2$ increases since $\sigma_{aw}^2$ decreases in this case. Moreover, we conclude that for the case of $M>1$, the covert rate decreases as the channel estimation errors increase and the deception strategy can achieve a higher covert rate than the benchmark when the channel estimation error is severe.}

\subsection{Discussion for scenario 3}
In Scenario 3, same as Scenario 2, Alice's transmission can not be completely covert. We discuss how the detection threshold $\varepsilon $, channel gains and $M$ affect the covert transmission performance under statistical CSI.\par
\begin{figure}[!t]
	\centering
	\includegraphics[width=3.2in]{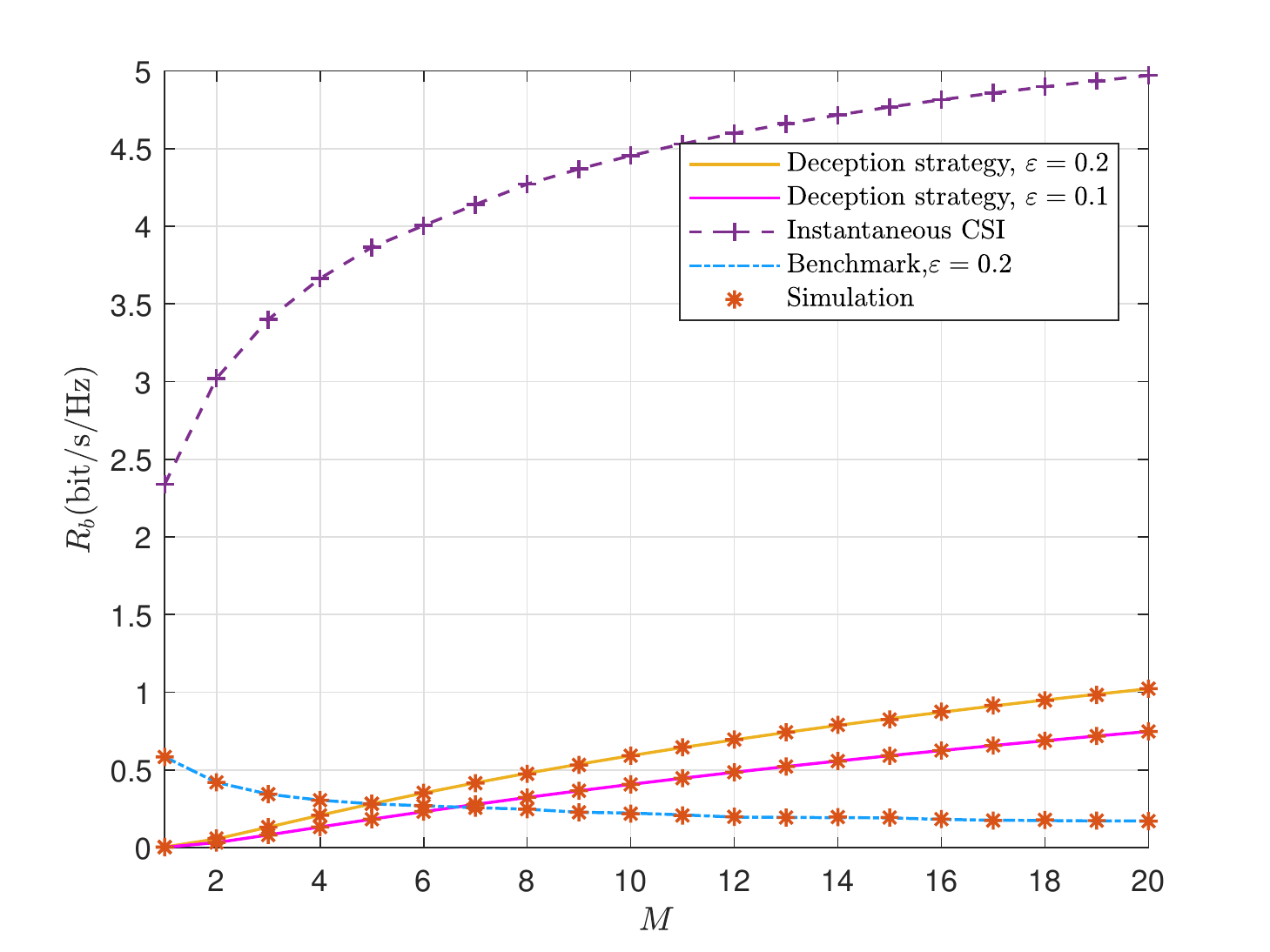}
	\caption{The covert rate versus $M$ with different channel gains under statistical CSI, $\sigma_{aw}^2=1$, $\sigma_{jw}^2=0.5$.}
	\label{fig_7}
\end{figure}
In Fig. \ref{fig_7}, the covert rate for the deception strategy and the benchmark are plotted when the statistical CSI is available. {We can find that when $M>1$, as $M$ increases, the covert rate of the deception strategy increases while that of the benchmark decreases. Furthermore, compared to instantaneous and imperfect CSI case, for the statistical CSI case, the deception strategy outperform the benchmark with a smaller number of antennas. For the benchmark, the statistical CSI of $\mathbf{h}_{jw}$ and $\mathbf{h}_{aw}$ affect the covert constraint under $\mathcal{H}_0$ and $\mathcal{H}_1$, respectively. When $M$ increases, the negative effect of only knowing statistical CSI also increases, which greatly affects the performance of the benchmark. However, for the deception strategy, the power detected by Willie under $\mathcal{H}_0$ and $\mathcal{H}_1$ are both mainly affected by the statistical CSI of $\mathbf{h}_{jw}$, which greatly reduces the impact of only knowing statistical CSI on performance. Therefore, when $M$ increases, the covert rate increases slightly. This behavior verifies our analysis in the \emph{Discussion} of Section \uppercase\expandafter{\romannumeral3}.C. When $M$ is quite small, same as the instantaneous and imperfect CSI case, due to the power constraints on $P_{j,1}$, the covert rate of the benchmark is greater than that of the deception strategy. However, As $M$ increases, the deception strategy increases in terms of covert rate while the benchmark decreases.} For comparison, the covert rate under instantaneous CSI is also illustrated in Fig. \ref{fig_7}. We can see that even if $\varepsilon = 0.2$, the covert rate under statistical CSI is much worse than that under instantaneous CSI, which is consistent with our intuition. Hence, obtaining the instantaneous CSI is an important and effective way to improve the covert rate in practice. \par
%In addition, for benchmark, we find the different performance under imperfect CSI and statistical CSI by comparing the Fig. \ref{fig_5} and the Fig. \ref{fig_7}. I It is because the increase of $M$ has positive effect under the imperfect CSI while it has negative effect under the statistical CSI.
%5\begin{figure}[!t]
%	\centering
%	\includegraphics[width=3.0in]{7_statistical_CSI_R_update.eps}
%	\caption{The covert rate versus $M$ with different channel gains under statistical CSI ($\sigma_{aw}^2<\sigma_{jw}^2$).}
%	\label{fig_8}
%\end{figure}
%The covert rate under statistical CSI are shown in Fig. \ref{fig_8} when $\sigma_{aw}^2<\sigma_{jw}^2$. We can find that for the deception strategy, different from Fig. \ref{fig_7}, the covert rate increases dramatically with the increase of $M$. Willie's detection of Alice's transmission is weaker in this case, allowing Alice to transmit with a higher power, and Jammer's deception worked effectively. However, for the benchmark, similar to the case of $\sigma_{aw}^2>\sigma_{jw}^2$, the covert rate decreases with the increase of $M$. Combining the analyses of Fig. \ref{fig_7} and Fig. \ref{fig_8}, we find that the deception strategy has much more advantages than the benchmark under statistical CSI.
\begin{figure}[!t]
	\centering
	\includegraphics[width=3.2in]{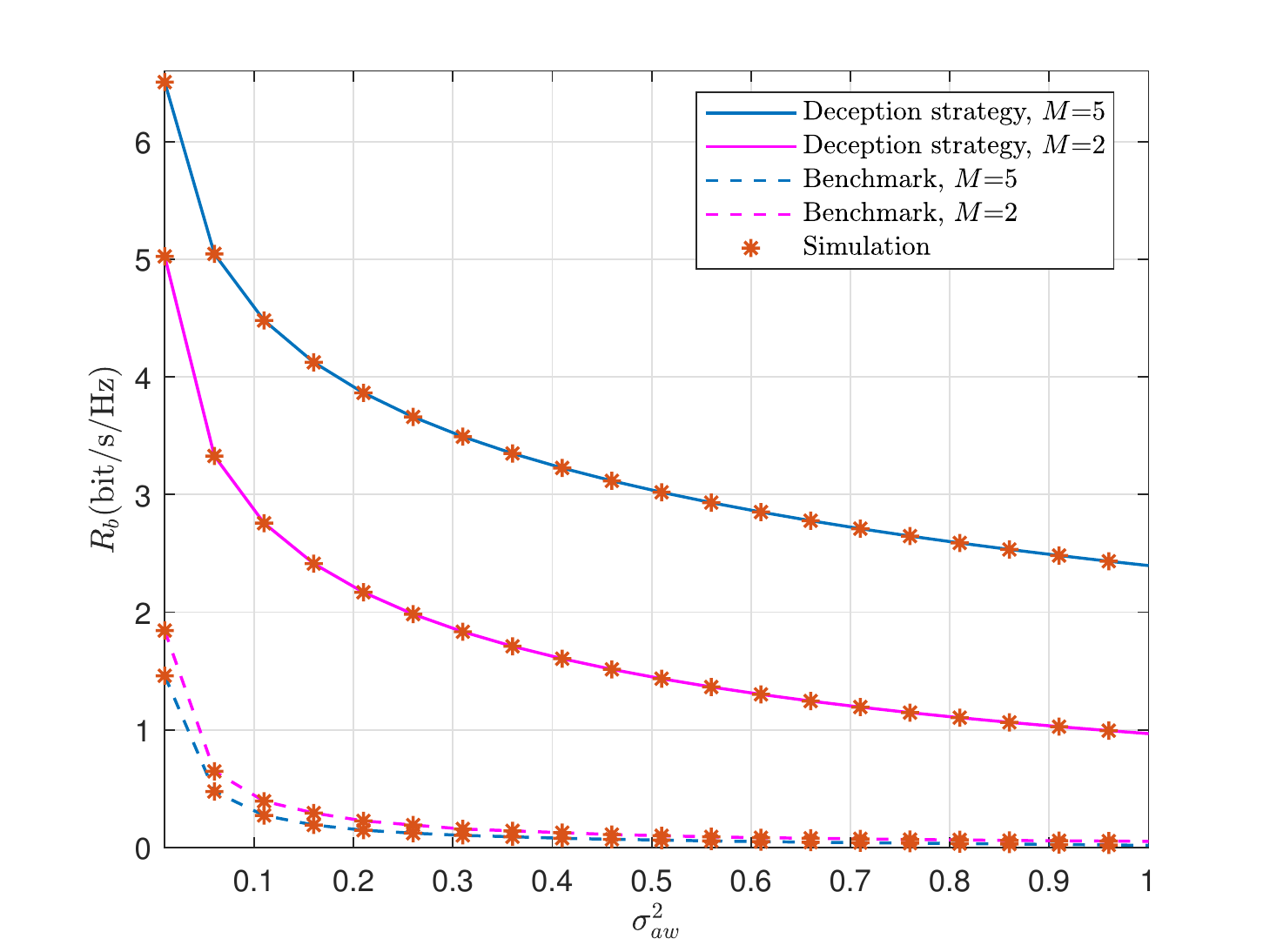}
	\caption{The covert rate versus $\sigma_{aw}^2$ with different $M$ under statistical CSI, $\sigma_{jw}^2=0.5$, $\varepsilon=0.2$.}
	\label{fig_8}
\end{figure}
{For the case when only statistical CSI is available, Fig. \ref{fig_8} shows the covert rate of the deception strategy and the benchmark versus $\sigma_{aw}^2$. We can find that, when $\sigma_{aw}^2$ increases, for the same $M$, both the performance of the deception strategy and that of the benchmark become worse. This is because Willie's detection of Alice's transmission is stronger in this case, which greatly reduces the transmission power of Alice. Moreover, for the same $\sigma_{aw}^2$,  $M$ has a positive effect on the covert rate with the deception strategy while has a negative effect on the covert rate with the benchmark, which matches the behavior shown in Fig. \ref{fig_7}.}

\section{Conclusion}
In this paper, we propose a cooperative deception strategy for covert transmission in presence of a multi-antenna adversary. Specifically, Jammer protects Alice's covert transmission by sending jamming signals under both ${\mathcal{H}_0}$ and ${\mathcal{H}_1}$ to attract Willie. Under three different CSI scenarios, $P_a$, $P_{j,0}$ and $P_{j,1}$ are jointly optimized to maximize the covert rate subject to the covert and total power constraints. Numerical results show that for the proposed deception strategy, the increase of the number of antennas has a positive effect on  the covert rate under three different CSI scenarios. {Additionally, compared to the benchmark, our proposed strategy is more robust with severe estimation errors.}

%\section*{Acknowledgments}
%This should be a simple paragraph before the References to thank those %individuals and institutions who have supported your work on this article.

{\appendices
%Use $\backslash${\tt{appendix}} if you have a single appendix:
%Do not use $\backslash${\tt{section}} anymore after %$\backslash${\tt{appendix}}, only $\backslash${\tt{section*}}.
%If you have multiple appendixes use $\backslash${\tt{appendices}} then use $\backslash${\tt{section}} to start each appendix.
%You must declare a $\backslash${\tt{section}} before using any $\backslash${\tt{subsection}} or using $\backslash${\tt{label}} ($\backslash${\tt{appendices}} by itself
\section{Derivation of $\mathbb{E}\left\lbrace D\left ( \mathbb{P}_{0}\|\mathbb{P}_{1} \right  )\right\rbrace$}
Let $b=\frac{\sigma _{aw}^{2}}{\sigma _{jw}^{2}}$, \eqref{eq:50} can be split into three parts, and $ \mathbb{E}\left\lbrace D\left ( \mathbb{P}_{0}\|\mathbb{P}_{1} \right  )\right\rbrace = Mb^M\left(A+B+C \right)$, where
\begin{subequations}\label{eq:74}
	\begin{align}
		A=&\int_{0}^{\infty}\mathrm{ln}\frac{\alpha z+\gamma }{\beta }\left(b+z \right)^{-\left ( M+1 \right )}dz,\label{eq:74a}\\
		B=&\int_{0}^{\infty}\frac{\beta}{\alpha z +\gamma}\left(b+z \right)^{-\left ( M+1 \right )}dz, \ \label{eq:74b}\\
		C=&-\int_{0}^{\infty}\left(b+z \right)^{-\left ( M+1 \right )}dz. \label{eq:74c}
	\end{align}
\end{subequations}
Integrating \eqref{eq:74a} by parts, we can get
\begin{equation}\label{eq:75}
	\begin{split}
		A=&-\frac{1}{M}\int_{0}^{\infty}\mathrm{ln}\frac{\alpha z+\gamma }{\beta }d\left(b+z \right)^{-M}\\
		=&-\frac{1}{M}\mathrm{ln}\frac{\alpha z+\gamma }{\beta }\left(b+z \right)^{-M}\big|_{0}^{\infty}\\
		&+\frac{1}{M}\int_{0}^{\infty}\frac{1}{z+\frac{\gamma}{\alpha }}\left(b+z \right)^{-M}dz.
	\end{split}
\end{equation}
With the help of the following identity \cite{2007247}
%[30] »ý·Ö±í317 Table of Integrals, Series, and Products Seventh Edition
\begin{equation}\label{eq:76}
	\begin{split}
		&\int_{0}^{\infty}x^{\nu -1}\left ( \beta +x \right 	)^{-\mu}\left ( \gamma +x \right )^{-\varrho }dx\\
		=&\beta ^{-\gamma}\beta ^{\nu-\varrho}\mathrm{B}\left ( \nu,\mu-\nu+\varrho \right )_{2}F_{1}\left ( \mu,\nu;\mu+\varrho ; 1-\frac{\gamma}{\beta} \right ),\\
		&\left [ \left | \mathrm{arg}\beta \right |<\pi, \left | \mathrm{arg}\gamma \right |<\pi, \mathrm{Re}\ \nu>0,  \mathrm{Re}\ \mu>\mathrm{Re}\ \left ( \nu-\varrho \right )\right ]	
	\end{split}
\end{equation}
After some algebraic operations, \eqref{eq:75} can be expressed as
\begin{equation}\label{eq:77}
	\begin{split}
		A=&M^{-1}b^{-M}\mathrm{ln}\frac{\gamma}{\beta}+\frac{\alpha}{\gamma}M^{-1}b^{1-M}\mathrm{B}\left ( 1,M \right )\\
		&\times _{2}F_{1}\left ( 1,1;1+M ; 1-\frac{b\alpha}{\gamma} \right )\\
		=&M^{-1}b^{-M}\mathrm{ln}\frac{\gamma}{\beta}\\
		&+\frac{\alpha}{\gamma}M^{-2}{b^{1-M}}_{2}F_{1}\left ( 1,1;1+M ; 1-\frac{b\alpha}{\gamma} \right ),
	\end{split}
\end{equation}
where $_{2}F_{1}\left (\alpha, \beta; \gamma; z \right )$ is the Hypergeometric Function.
Similarly, \eqref{eq:74b} can be expressed as follow
\begin{equation}\label{eq:78}
	\begin{split}
		B=&\frac{\beta}{\alpha}\int_{0}^{\infty}\frac{1}{z+\frac{\gamma}{\alpha }}\left(b+z \right)^{-M-1}dz\\
		=&\frac{\beta}{\alpha}\frac{\alpha}{\gamma}b^{-M}\mathrm{B}\left ( 1,M+1 \right )_{2}F_{1}\left ( 1,1;2+M ; 1-\frac{b\alpha}{\gamma} \right )\\
		=&\frac{\beta }{\gamma }\left(M+1\right)  ^{-1}{b^{-M}}_{2}F_{1}\left ( 1,1;2+M ; 1-\frac{b\alpha}{\gamma} \right ).
	\end{split}
\end{equation}
In addition, after some operations, the final expression of \eqref{eq:74c} can be obtained as
\begin{equation}\label{eq:79}
	C=\frac{1}{M}\left(b+z \right)^{-M}\big|_{0}^{\infty}=-M^{-1}b^{-M}.
\end{equation}

\section{Derivation of $\frac{\partial }{\partial \gamma }\mathbb{E}\left\lbrace D\left ( \mathbb{P}_{0}\|\mathbb{P}_{1} \right  )\right\rbrace$}
Substituting (51) into  $\frac{\partial }{\partial \gamma }\mathbb{E}\left\lbrace D\left ( \mathbb{P}_{0}\|\mathbb{P}_{1} \right  )\right\rbrace$, we get
\begin{equation}\label{eq:80}
	\begin{split}
		&\frac{\partial }{\partial \gamma }\mathbb{E}\left\lbrace D\left ( \mathbb{P}_{0}\|\mathbb{P}_{1} \right  )\right\rbrace\\
		=&\frac{1}{\gamma }-\frac{\alpha}{\gamma^2 }bM^{-1}{_2F_1}\left ( 1,1;1+M;\frac{\gamma - \alpha b}{\gamma} \right )\\
		&+\frac{\alpha}{\gamma }bM^{-1}\left \{ \frac{\partial }{\partial \gamma}\left [{_2F_1}\left ( 1,1;1+M;\frac{\gamma - \alpha b}{\gamma} \right )  \right ] \right \}\\
		&-\frac{\beta}{\gamma^2 }b\frac{M}{M+1}{_2F_1}\left ( 1,1;2+M;\frac{\gamma - \alpha b}{\gamma}\right )\\
		&+\frac{\beta}{\gamma }b\frac{M}{M+1}\left \{ \frac{\partial }{\partial \gamma }\left [{_2F_1}\left ( 1,1;2+M;\frac{\gamma - \alpha b}{\gamma} \right )  \right ] \right \},
	\end{split}
\end{equation}
where ${_2F_1}\left ( \alpha, \beta;\gamma;z \right )$ is expressed as\cite{2007859}
\begin{equation}\label{eq:81}
	\begin{split}
		\frac{1}{\mathrm{B}\left ( \beta, \gamma-\beta \right )}\int_{0}^{1}t^{\beta -1}\left ( 1-\right.&\left. t \right )^{\gamma-\beta-1}\left ( 1-tz \right )^{-\alpha}dt,\\
		&\left [ \mathrm{Re}\ \gamma>\mathrm{Re}\ \beta> 0\right ].
	\end{split}
\end{equation}
Thus, the second term of \eqref{eq:80} is given by
\begin{equation}\label{eq:82}
	\begin{split}
		&\frac{\partial}{\partial \gamma}\left\lbrace {_2F_1}\left ( 1,1;1+M;\frac{\gamma - \alpha b}{\gamma} \right )\right\rbrace \\
		=&\frac{\partial}{\partial \gamma}\left\lbrace M\int_0^{1} \left ( 1-t \right )^{M-1}\left ( 1-\frac{\gamma -\alpha b}{\gamma }t \right )^{-1}dt\right\rbrace\\
		=&\frac{\alpha b}{\gamma ^2}\left(M+1 \right) ^{-1} {_2F_1}\left ( 2,2;2+M;\frac{\gamma - \alpha b}{\gamma} \right ).
	\end{split}
\end{equation}\par
The derivation of the fourth term of \eqref{eq:80} is similar to \eqref{eq:82}. Finally, the expression of $\frac{\partial }{\partial \gamma }\mathbb{E}\left\lbrace D\left ( \mathbb{P}_{0}\|\mathbb{P}_{1} \right  )\right\rbrace$ can be obtained.
}


%\section{References Section}
%You can use a bibliography generated by BibTeX as a .bbl file.
% The IEEEtran BibTeX style support page is:

%\section{Simple References}
%You can manually copy in the resultant .bbl file and set second argument of $\backslash${\tt{begin}} to the number of references
% (used to reserve space for the reference number labels box).

\bibliographystyle{IEEEtran}
\bibliography{ref}

\begin{thebibliography}{10}
\providecommand{\url}[1]{#1}
\csname url@samestyle\endcsname
\providecommand{\newblock}{\relax}
\providecommand{\bibinfo}[2]{#2}
\providecommand{\BIBentrySTDinterwordspacing}{\spaceskip=0pt\relax}
\providecommand{\BIBentryALTinterwordstretchfactor}{4}
\providecommand{\BIBentryALTinterwordspacing}{\spaceskip=\fontdimen2\font plus
\BIBentryALTinterwordstretchfactor\fontdimen3\font minus
  \fontdimen4\font\relax}
\providecommand{\BIBforeignlanguage}[2]{{%
\expandafter\ifx\csname l@#1\endcsname\relax
\typeout{** WARNING: IEEEtran.bst: No hyphenation pattern has been}%
\typeout{** loaded for the language `#1'. Using the pattern for}%
\typeout{** the default language instead.}%
\else
\language=\csname l@#1\endcsname
\fi
#2}}
\providecommand{\BIBdecl}{\relax}
\BIBdecl

\bibitem{8438892}
Y.~Zhang, Y.~Shen, X.~Jiang, and S.~Kasahara, ``Mode selection and spectrum
  partition for {D2D} inband communications: A physical layer security
  perspective,'' \emph{IEEE Trans. Commun.}, vol.~67, no.~1, pp. 623--638,
  2019.

\bibitem{9205225}
M.~Letafati, A.~Kuhestani, K.-K. Wong, and M.~J. Piran, ``A lightweight secure
  and resilient transmission scheme for the internet of things in the presence
  of a hostile jammer,'' \emph{IEEE Internet of Things J.}, vol.~8, no.~6, pp.
  4373--4388, 2021.

\bibitem{7514758}
Y.~Huang, J.~Wang, C.~Zhong, T.~Q. Duong, and G.~K. Karagiannidis, ``Secure
  transmission in cooperative relaying networks with multiple antennas,''
  \emph{IEEE Trans. Wireless Commun.}, vol.~15, no.~10, pp. 6843--6856, 2016.

\bibitem{8714018}
S.~Yan, Y.~Cong, S.~V. Hanly, and X.~Zhou, ``Gaussian signalling for covert
  communications,'' \emph{IEEE Trans. Wireless Commun.}, vol.~18, no.~7, pp.
  3542--3553, 2019.

\bibitem{6584948}
B.~A. Bash, D.~Goeckel, and D.~Towsley, ``Limits of reliable communication with
  low probability of detection on {AWGN} channels,'' \emph{IEEE J. Sel. Areas
  Commun.}, vol.~31, no.~9, pp. 1921--1930, 2013.

\bibitem{7407378}
M.~R. Bloch, ``Covert communication over noisy channels: A resolvability
  perspective,'' \emph{IEEE Trans. Inf. Theory}, vol.~62, no.~5, pp.
  2334--2354, 2016.

\bibitem{7084182}
S.~Lee, R.~J. Baxley, M.~A. Weitnauer, and B.~Walkenhorst, ``Achieving
  undetectable communication,'' \emph{IEEE J. Sel. Topics Signal Process.},
  vol.~9, no.~7, pp. 1195--1205, 2015.

\bibitem{7447769}
L.~Wang, G.~W. Wornell, and L.~Zheng, ``Fundamental limits of communication
  with low probability of detection,'' \emph{IEEE Trans. Inf. Theory}, vol.~62,
  no.~6, pp. 3493--3503, 2016.

\bibitem{7805182}
B.~He, S.~Yan, X.~Zhou, and V.~K.~N. Lau, ``On covert communication with noise
  uncertainty,'' \emph{IEEE Commun. Lett.}, vol.~21, no.~4, pp. 941--944, 2017.

\bibitem{7352320}
D.~Goeckel, B.~Bash, S.~Guha, and D.~Towsley, ``Covert communications when the
  warden does not know the background noise power,'' \emph{IEEE Commun. Lett.},
  vol.~20, no.~2, pp. 236--239, 2016.

\bibitem{8761935}
H.~Q. Ta and S.~W. Kim, ``Covert communication under channel uncertainty and
  noise uncertainty,'' in \emph{ICC 2019}, 2019, pp. 1--6.

\bibitem{8471218}
J.~Wang, W.~Tang, Q.~Zhu, X.~Li, H.~Rao, and S.~Li, ``Covert communication with
  the help of relay and channel uncertainty,'' \emph{IEEE Wireless Commun.
  Lett.}, vol.~8, no.~1, pp. 317--320, 2019.

\bibitem{6970786}
P.~H. Che, M.~Bakshi, C.~Chan, and S.~Jaggi, ``Reliable deniable communication
  with channel uncertainty,'' in \emph{2014 IEEE Inf. Theory Workshop}, 2014,
  pp. 30--34.

\bibitem{8379465}
S.~Yan, B.~He, X.~Zhou, Y.~Cong, and A.~L. Swindlehurst, ``Delay-intolerant
  covert communications with either fixed or random transmit power,''
  \emph{IEEE Trans. Inf. Forensics Security}, vol.~14, no.~1, pp. 129--140,
  2019.

\bibitem{7579596}
B.~A. Bash, D.~Goeckel, and D.~Towsley, ``Covert communication gains from
  adversary’s ignorance of transmission time,'' \emph{IEEE Trans. Wireless
  Commun.}, vol.~15, no.~12, pp. 8394--8405, 2016.

\bibitem{7964713}
T.~V. Sobers, B.~A. Bash, S.~Guha, D.~Towsley, and D.~Goeckel, ``Covert
  communication in the presence of an uninformed jammer,'' \emph{IEEE Trans.
  Wireless Commun.}, vol.~16, no.~9, pp. 6193--6206, 2017.

\bibitem{7421206}
T.~V. Sobers, B.~A. Bash, D.~Goeckel, S.~Guha, and D.~Towsley, ``Covert
  communication with the help of an uninformed jammer achieves positive rate,''
  in \emph{2015 49th Asilomar Conf. Signals, Systems and Computers}, 2015, pp.
  625--629.

\bibitem{8445707}
R.~Soltani, D.~Goeckel, D.~Towsley, B.~A. Bash, and S.~Guha, ``Covert wireless
  communication with artificial noise generation,'' \emph{IEEE Trans. Wireless
  Commun.}, vol.~17, no.~11, pp. 7252--7267, 2018.

\bibitem{9456866}
R.~Sun, B.~Yang, S.~Ma, Y.~Shen, and X.~Jiang, ``Covert rate maximization in
  wireless full-duplex relaying systems with power control,'' \emph{IEEE Trans.
  Commun.}, vol.~69, no.~9, pp. 6198--6212, 2021.

\bibitem{8519751}
K.~Shahzad, X.~Zhou, S.~Yan, J.~Hu, F.~Shu, and J.~Li, ``Achieving covert
  wireless communications using a full-duplex receiver,'' \emph{IEEE Trans.
  Wireless Commun.}, vol.~17, no.~12, pp. 8517--8530, 2018.

\bibitem{9398675}
S.~Ma, Y.~Zhang, H.~Li, S.~Lu, N.~Al-Dhahir, S.~Zhang, and S.~Li, ``Robust
  beamforming design for covert communications,'' \emph{IEEE Trans. Inf.
  Forensics Security}, vol.~16, pp. 3026--3038, 2021.

\bibitem{9685518}
X.~Peng, J.~Wang, S.~Xiao, and W.~Tang, ``Strategies in covert communication
  with imperfect channel state information,'' in \emph{2021 GLOBECOM}, 2021,
  pp. 1--6.

\bibitem{9390203}
C.~Wu, S.~Yan, X.~Zhou, R.~Chen, and J.~Sun, ``Intelligent reflecting surface
  {(IRS)} -aided covert communication with warden’s statistical {CSI},''
  \emph{IEEE Wireless Commun. Lett.}, vol.~10, no.~7, pp. 1449--1453, 2021.

\bibitem{9438645}
J.~Si, Z.~Li, Y.~Zhao, J.~Cheng, L.~Guan, J.~Shi, and N.~Al-Dhahir, ``Covert
  transmission assisted by intelligent reflecting surface,'' \emph{IEEE Trans.
  Commun.}, vol.~69, no.~8, pp. 5394--5408, 2021.

\bibitem{9381893}
O.~Shmuel, A.~Cohen, and O.~Gurewitz, ``Multi-antenna jamming in covert
  communication,'' \emph{IEEE Trans. Commun.}, vol.~69, no.~7, pp. 4644--4658,
  2021.

\bibitem{8878022}
K.~Shahzad, X.~Zhou, and S.~Yan, ``Covert wireless communication in presence of
  a multi-antenna adversary and delay constraints,'' \emph{IEEE Trans. Veh.
  Technol.}, vol.~68, no.~12, pp. 12\,432--12\,436, 2019.

\bibitem{2005Testing}
E.~L. Lehmann, ``Testing statistical hypotheses,'' \emph{Publications of the
  American Statistical Association}, vol. 101, no. 474, pp. 847--848, 2005.

\bibitem{9122034}
J.~Si, Z.~Cheng, Z.~Li, J.~Cheng, H.-M. Wang, and N.~Al-Dhahir, ``Cooperative
  jamming for secure transmission with both active and passive eavesdroppers,''
  \emph{IEEE Trans. Commun.}, vol.~68, no.~9, pp. 5764--5777, 2020.

\bibitem{2006Elements}
T.~M. Cover and J.~A. Thomas, ``Elements of information theory (2. ed.),''
  \emph{Tsinghua University Pres}, 2006.

\bibitem{9634882}
J.~Bai, J.~He, X.~Jiang, and L.~Chen, ``Performance analysis for dual-hop
  covert communication system with outdated {CSI},'' in \emph{2021
  International Conf. NaNA}, 2021, pp. 206--211.

\bibitem{2017A}
J.~G. Vanantwerp and R.~D. Braatz, ``A tutorial on linear and bilinear matrix
  inequalities,'' \emph{J. Process Control}, vol.~10, no.~4, pp. 363--385,
  2017.

\bibitem{2007247}
``3–4 - definite integrals of elementary functions,'' in \emph{Table of
  Integrals, Series, and Products (Seventh Edition)}, seventh edition~ed.,
  A.~Jeffrey, D.~Zwillinger, I.~Gradshteyn, and I.~Ryzhik, Eds.\hskip 1em plus
  0.5em minus 0.4em\relax Boston: Academic Press, 2007, pp. 247--617.

\bibitem{2007859}
``8–9 - special functions,'' in \emph{Table of Integrals, Series, and
  Products (Seventh Edition)}, seventh edition~ed., A.~Jeffrey, D.~Zwillinger,
  I.~Gradshteyn, and I.~Ryzhik, Eds.\hskip 1em plus 0.5em minus 0.4em\relax
  Boston: Academic Press, 2007, pp. 859--1048.

\end{thebibliography}

% your own custom macro containing the $\backslash${\tt{includegraphics}} command to make things


%\bf{If you include a photo:}\vspace{-33pt}
%\begin{IEEEbiography}[{\includegraphics[width=1in,height=1.25in,clip,keepaspectratio]{fig%1}}]{Michael Shell}
%Use $\backslash${\tt{begin\{IEEEbiography\}}} and then for the 1st argument use $\backslash${\tt{includegraphics}} to declare and link the author photo.
%Use the author name as the 3rd argument followed by the biography text.
%\end{IEEEbiography}



%\vfill

\end{document}